\documentclass[aps,prb,twocolumn,superscriptaddress,nofootinbib]{revtex4}

\usepackage{graphics}
\usepackage{amssymb}
\usepackage{graphicx}
\usepackage{epsf,epsfig}
\usepackage{amsfonts}
\usepackage{amsmath}
\usepackage{color}

\newcommand{\eps}{\varepsilon}
\newcommand{\bfq}{{\bf q}}
\newcommand{\bfk}{{\bf k}}

\begin{document}

\title{Transport in topological insulators with bulk-surface coupling: 
Interference corrections and conductance fluctuations}

\author{H. Velkov}

\affiliation{Institut f\"ur Physik, Johannes Gutenberg Universit\"at Mainz, 55128 Mainz, Germany}
\affiliation{Spin Phenomena Interdisciplinary Center (SPICE)}

\author{G.~N.~Bremm}
\affiliation{Centro Brasileiro de Pesquisas F\'{i}sicas, Rua Xavier Sigaud 150, 22290-180, Rio de Janeiro, Brazil
}

\author{T. Micklitz}
\affiliation{Centro Brasileiro de Pesquisas F\'{i}sicas, Rua Xavier Sigaud 150, 22290-180, Rio de Janeiro, Brazil
}

\author{G. Schwiete}
\email{gschwiete@ua.edu}
\affiliation{Department of Physics and Astronomy, Center for Materials for Information Technology (MINT), The University of Alabama, Alabama 35487, USA}

\date{\today}

\begin{abstract}
Motivated by the experimental difficulty to produce topological insulators (TIs) of the 
Bi$_2$Se$_3$ family with pure surface-state conduction, we study the effect that the 
bulk can have on the low-temperature transport properties of gated thin TI films. 
In particular, we focus on interference corrections, namely weak localization (WL) or 
weak-antilocalization (WAL), and conductance fluctuations (CFs) based 
on an effective low-energy Hamiltonian. Utilizing diagrammatic perturbation theory 
we first analyze the bulk and the surface separately and subsequently discuss WL/WAL 
and CFs when a tunneling-coupling is introduced. We identify the relevant soft diffusion 
modes of the coupled system and use this insight to make simultaneous predictions for 
both interference corrections and conductance fluctuations in various parameter regimes. 
The results strongly suggest that the combined measurement of both quantities can provide an 
improved understanding of the physics underlying the low-temperature transport 
processes in thin TI films with bulk-surface coupling.
\end{abstract}

\maketitle

\section{Introduction}

During the last decade, the field of condensed-matter physics witnessed the 
advent of a new class of materials, the three-dimensional (3D) topological insulators (TIs). 
First discussed theoretically,~\cite{Fu07} their existence and peculiar band structure has been 
verified experimentally by ARPES measurements.~\cite{Hsieh08,Xia09,Zhang09,Hasan10} 3D TIs 
provide conduction along their surfaces, while ideally their bulk is insulating.~\cite{Hasan10,Qi11} 
The experimental observation of a pure surface conduction, however, proved to be a formidable task. 
Indeed, a perfectly insulating bulk is difficult to obtain for the well-studied topological insulator materials 
$Bi_2Se_3$ and $Bi_2Te_3$. These TIs have narrow bulk band gaps, so that accidental impurities or vacancies 
tend to populate the bulk conduction band. 
Electrical gating, using thin film samples,~\cite{Chen10,Checkelsky11} doping and 
annealing~\cite{Analytis10,Checkelsky09,Qu10} can help reducing the bulk electron density, 
but do not remove it completely.  
The presence of bulk conduction electrons and their coupling to the surface have profound 
consequences for the low-temperature transport properties of the 3D topological insulators. 

Specifically, quantum interference corrections to the Drude conductivity 
are sensitive to the presence of conducting bulk states. 
3D TI materials with a perfectly insulating bulk are expected to exhibit 
a `single-channel' weak anti-localization (WAL),~\cite{Hikami80}
i.e. a reduction of the conductivity in response to 
a weak (perpendicular) magnetic field suppressing quantum interference effects. 
`Single channel' here refers to a single Dirac cone of surface states and
the positive magnetoresistance
is a consequence of strong spin-orbit coupling in the conducting surfaces. WAL
has been observed in the majority of experiments performed on thin TI films.~\cite{Chen10,Checkelsky11,Steinberg11,Kim13,Chiatti16} 
 Still, the interpretation of experimental data is not always straightforward.  
Ref.~\onlinecite{Steinberg11}, e.g., reports on WAL changing from a `one-' 
to a `two-channel contribution'  
as the gate voltage is varied, attributed to a change in the relation between bulk-surface scattering time and the phase-coherence time. Ref.~\onlinecite{Kim13} observes a similar crossover in very thin films, which is here explained by a change in direct inter-surface tunneling.  
Further experiments report on a crossover from a positive to a negative magnetoresistance,
or weak localization (WL), in ultra-thin samples (with thickness about 5 nm) when the applied parallel~\cite{Wang14} or 
perpendicular~\cite{Zhang13} magnetic field exceeds a respective threshold.
Finally, experiments on the TI material $Bi_2Te_2Se$ reported in Ref.~\onlinecite{Li91} 
stressed the importance of a bulk-mediated coupling between opposite surface states. 

There have been several works on the conductivity of topological insulators 
under non-ideal situations, 
studying, e.g., effects of magnetic doping\cite{Lu11} and spin-orbit coupled impurities,\cite{Shan12,Adroguer15} and contributions from bulk channels.\cite{Lu11a,Shan12,Ostrovsky12} We here build on the theoretical 
work by Garate and Glazman~\cite{Garate12}, 
and include conductance fluctuations into the picture.
In their 
work, Garate and Glazman 
demonstrated that once bulk states are included into the picture 
 the conductivity of TIs shows a rich  
 behavior beyond the single-channel WAL.~\cite{Garate12}
 Building on a model that accounts for both, bulk and 
surface states and their coupling, 
they identified several different regimes exhibiting WL or WAL and with varying number 
of contributing channels. 
As already discussed, similar behavior has been observed in experiment; 
however, a detailed comparison with experiment can be complicated by the uncertainty in the system parameters. 
In this situation, the study of a further independent observable should be of help. Conductance fluctuations (CFs) 
can be measured in parallel with the conductance. Examples of such measurements on different TI materials include $Bi_2Se_3Se$,\cite{Matsuo12,Matsuo13} $Bi_2Te_2Se$,\cite{Li12,Li14} $Bi_2Se_3$,\cite{Kim13,Kandala13} $Bi{}_{1.5}Sb{}_{0.5}Te{}_{1.8}Se{}_{1.2}$,\cite{Xia13} and $Bi_2Te_2S$.\cite{Trivedi16} In experiment, CFs become visible through aperiodic patterns in the magnetoresistance. 
Similar as the interference corrections to conductivity, CFs are expected to depend sensitively on the presence 
of conducting bulk states. 
In this paper, we employ the model of Garate and Glazman to calculate CFs for 
 thin TI films of the $Bi_2Se_3$ family. 
  When combined with the results for the conductivity, 
we arrive at a set of simultaneous predictions for WL/WAL corrections and CFs 
for the different characteristic transport regimes.

The paper is structured as follows. 
Sec.~\ref{sec:results} introduces the model and summarizes main results of this work. 
The following sections elaborate on the results, and present technical details of the calculation. 
In Sec.~\ref{sec:soft_modes} we discuss the soft diffusion modes that govern 
the low-temperature transport in the system. We first analyze bulk and surface soft modes 
separately and then discuss a finite bulk-surface-coupling. 
Sections \ref{sec:Interference_corrections}  and ~\ref{sec:conductance_fluctuations}  
are devoted to, respectively, the calculation of WL/WAL corrections to the conductivity and  
CFs. We conclude in section~\ref{sec:conclusion}. 
Several appendices cover further details of the calculations presented in the main text.

\section{Model and Results}
\label{sec:results}

\subsection{Model}
\label{subsec:model}
To describe the clean 3D TI thin film we start out from the 
Hamiltonian,
\begin{align}
\label{H0}
\hat H_0=
\hat H_{\text{bulk}}+	\hat H_{\text{surf.}}+\hat H_{\text{coupl.}}.
\end{align}
Here the bulk- and surface-Hamiltonian have the standard Dirac form,
\begin{align}
	\hat H_{\text{bulk}}
	&= 
	M \mathbf{1}_{2} \otimes \tau_z 
	+ 
	\hbar v_B \mathbf{k} 
	\cdot 
	\boldsymbol\sigma \otimes \tau_x,
\label{intro_bulk_Hamiltonian}      
\\
	\hat H_{\text{surf.}}
	&= 
	\hbar v_S \mathbf{k}_\parallel 
	\cdot 
	\boldsymbol\sigma,
\label{intro_surface_Hamiltonian}      
\end{align}
 with $\mathbf{k}_\parallel=(k_x,k_y)$, $\bold{k}=(\mathbf{k}_\parallel,k_z)$ surface- and 
 bulk-momenta, respectively, and
  Pauli matrices $\tau_i$ and $\sigma_i$ 
operating in the space of orbitals and spin, 
and $i=x,y,z$. 
The bulk-Hamiltonian, $\hat H_{\text{bulk}}$, 
is in the strong topological phase which guarantees the 
presence of a single Dirac cone, 
$\hat H_{\text{surf.}}$,
at its surfaces. 
In the following we assume that the TI is electrically gated on one of its surfaces. 
The corresponding surface states are separated from the bulk, and 
 the tunneling matrix, $\hat H_{\text{coupl.}}$, accounts 
for a local coupling between the latter and bulk states.
The coupling is due to (random) tunneling processes which neither conserve the spin or momentum of the charge carrier. It is modeled 
 by a random matrix drawn from the symplectic ensemble, as discussed 
in more detail in Sec.~\ref{SMcoupled}.

The above Hamiltonian~\eqref{H0}, e.g., provides an effective low-energy description of
 a clean gated topological insulator thin film from the $Bi_2Se_3$ family.~\cite{Zhang09}
The bulk spectrum is that of massive Dirac electrons with mass-gap $M$,
\begin{equation}
	 \epsilon_{\mathbf{k}\pm} = \pm \sqrt{\hbar ^2 v_B^2 k^2 + M^2},
\label{bulk_energy_spectrum}
\end{equation}
where the index $\pm$ distinguishes between the two bands of Kramer's degenerate states.
Relevant to us is the conduction band, spanned by the eigenspinors 
\begin{align}
	\label{bulk:eigenvectors}
	  \left| 1, \mathbf{k} \right> 
	  &= \sqrt{\tfrac{\epsilon_k + M}{2 \epsilon_k}}
	  \left ( 1, 0, \tfrac{\hbar v_B k_z}{\epsilon_k + M}, \tfrac{\hbar v_B k_+}{\epsilon_k + M} \right),
	\nonumber \\
	  \left| 2, \mathbf{k} \right> &= \sqrt{\tfrac{\epsilon_k + M}{2 \epsilon_k}}
	  \left ( 0, 1, \tfrac{\hbar v_B k_-}{\epsilon_k + M}, \tfrac{-\hbar  v_B k_z}{\epsilon_k + M} \right).
\end{align}
Here $k_\pm = k_x \pm i k_y$,   
and the components are given in the 
$\tau_z\otimes\sigma_z$-eigenbasis which in the following are denoted by 
$\left\{ \left| T\uparrow \right>, \left| T\downarrow \right>, \left| B\uparrow \right>, \left| B\downarrow \right> \right\}$.
The twofold degeneracy of bulk-bands reflects the presence of time-reversal and inversion symmetries in the model. 
The spectrum of surface-states is
$\epsilon_{\alpha}(\bold{k}) = \alpha \hbar v k$,
and the corresponding eigenspinors read
\begin{align}
        \left| \alpha, \mathbf{k} \right> 
	 &= \frac{1}{\sqrt{2}} (   \alpha e^{- i \phi_\bold{k}}, 1)^t,
	 \qquad \alpha=\pm,	\label{eq:SO_basis}
	\end{align}
with $\sin \phi_\bold{k}=k_y/k$, $\cos\phi_{\bf k}=k_x/k$. 
In the following we assume that the surface Fermi energy lies above the Dirac point, $\epsilon_{F,S}>0$, 
and introduce the simplified notation $|{\bf k}\rangle\equiv |+,{\bf k}\rangle$. 

Any realistic TI inevitably contains some degree of disorder, which can be accounted for by
\begin{align}
\label{H}
\hat H&= \hat H_0 + \hat V_{\text{dis}}.
\end{align} 
 We here restrict to weak disorder, which is uncorrelated for bulk and surface and does not couple 
to orbital or spin degrees of freedoms. 
That is, $\hat V_{\text{dis}}(\mathbf{r}) = V(\mathbf{r})\, \mathbf{1}_{4(2)}$ in the bulk (surface),
where $V$ is Gaussian distributed white noise 
characterized by the second moment 
$\langle  V(\mathbf{r}) V(\mathbf{r}')\rangle=u^{B(S)}_0\delta(\mathbf{r}-\mathbf{r}')$, and
weak disorder refers to $(\epsilon_{F,B}-M) \tau_{0B}\gg1$  ($ \eps_{F,S}\tau_{0S}\gg 1$). 
Notice that although we are considering isotropic disorder, the internal structure of model~\eqref{H0} implies 
a difference in the corresponding
elastic scattering and mean transport times, $\tau_{0B(S)}$ and $\tau_{B(S)}$. 
 
Physical properties of model~\eqref{H0} depend on 
the relative position of the bulk Fermi energy $\epsilon_F$ with respect to the gap $M$, 
and the strength of coupling between surface(s) and bulk states. The latter is defined by the ratio of tunneling rates and soft-mode masses, as discussed in detail in the next section. 
Both, the Fermi energy and the couplings 
are changed by modifying the electrical gate on the surface. Based on the Hamiltonian \eqref{H0} we will specifically discuss two models describing different physical realizations of TI films. Common to both models is the gate-induced depletion layer between one of the two surfaces, which we will refer to as S1, and the bulk. Correspondingly, the surface S1 and the bulk are coupled only through tunneling processes. For the \emph{TI thin film with one active surface}, the second surface (S2) is assumed to be weakly coupled and its diffusion modes gapped. A possible realization, suggested in Ref.~\onlinecite{Garate12}, is based on introducing random magnetic impurities in order to generate strong phase decoherence on surface S2. A second gate has also been proposed\cite{Garate12} as a means for diminishing the coupling between S2 and the bulk. The second model, the \emph{TI thin film with two active surfaces} is motivated by the observation that in the presence of only one gate and without further provisions, surface S2 can be expected to be well coupled to the bulk and to participate in transport at low temperatures. Within our approach, this situation can be modeled through a strong tunneling coupling between bulk and surface S2. Even in this limit, however, we will assume that the tunneling rate remains smaller than the disorder scattering rate.

\subsection{Main results}

We next summarize the main results 
 in the limits of small and large Fermi energies  
 and weak and strong coupling for TI films with one or two active surfaces. 
Concentrating on low temperatures, we assume that
 temperature broadening can be neglected and temperature dependencies originate primarily from inelastic scattering processes. 
General expressions covering regimes of intermediate coupling are provided in the subsequent sections of this paper and the enthusiastic reader can 
derive from these results in regimes of his interest. The role of temperature broadening is briefly discussed at the end of this summary in Sec.~\ref{sec:TD}.

Interference corrections to the magneto-conductance, $\Delta G(H)=G(H)-G(0)$, 
and conductance fluctuations, $\overline{\delta G^2}$,
are governed by the soft diffusion modes of the weakly disordered system. 
The identification of the soft modes in the different parameter regimes can provide us with 
a qualitative understanding of the main results. In this summary, we will discuss the magneto-conductance and CFs for system sizes much larger than all phase-coherence lengths. We will also comment on the opposite limit, in which phase-coherence lengths are large as compared to system sizes and the CFs become universal. 

We define the \emph{strong coupling limit} by the condition that the tunneling lengths in the system are much smaller than the phase coherence lengths. The opposite case will be called \emph{weak coupling limit}. In the limiting cases discussed below, the magneto-conductance and CFs in the strong coupling limit can be written in the form
\begin{align}
\Delta G(H)\simeq\alpha \frac{e^2}{\pi h} f(x_\varphi),\quad \overline{\delta G^2}\simeq \beta \left(\frac{e^2}{h}\right)^2 \frac{3}{2\pi}\left(\frac{l_\varphi}{L}\right)^2.\label{eq:ab1}
\end{align}
Here,  $f(z)=\ln(z)-\Psi(1/2+z)$ with $f(z)\approx\ln (z)$ 
 for $z\ll 1$ and $f(z)=- 1/(24z^2)$ 
 for $z\gg 1$. Moreover, $x_\varphi=\hbar/(4eHl^2_\varphi)$ and $l_\varphi$ is the effective phase coherence length in the strongly coupled system of lateral size $L$. In the weak coupling limit we further need to sum over separate surface and bulk contributions, which will be labeled by the letters $S$ and $B$. The coefficients $\alpha$ and $\beta$ encode information on the soft diffusion modes present in the regime of interest. The coefficient $\alpha$ derives from the number of relevant soft Cooperon modes. Each singlet Cooperon mode gives a contribution of $1/2$ to $\alpha$, while the contribution of each triplet Cooperon mode is $-1/2$. The coefficient $\beta$, in turn, simply counts the number of relevant soft diffusion modes, including both Cooperons and diffusons irrespective of their spin structure. In the universal limit, the result for the CF reads 
\begin{align}
\overline{\delta G^2}\simeq \gamma \beta (e^2/h)^2,\label{eq:ab2}
\end{align}
where $\gamma\simeq 0.093$ is a geometry-dependent factor and $\beta=\beta_S+\beta_B$ in the weak coupling limit.

\subsubsection{TI thin film with one active surface}

\paragraph{Strong coupling limit:}

In the strong coupling limit one obtains
\begin{align}
\alpha=1/2, \qquad \beta=2,
\end{align} 
independent of the position of the Fermi energy or the system-size $L$, 
provided that $L$ remains larger than the tunneling lengths. 
The value  $\beta=2$ for conductance fluctuations
indicates the presence of two soft modes, that is, 
each one diffuson- and one Cooperon-mode 
(see Sec.~\ref{sec:soft_modes} for details). 
The enhancement of conductivity (WAL) and negative magnetoresistance, $2\alpha=1$,  on the other hand,
 implies that the Cooperon mode, and by time-reversal symmetry also the diffuson mode, 
 is in the spin-singlet channel. 
That is, at strong coupling the system possesses
each one gapless spin-singlet mode in diffuson and Cooper channels, respectively, viz., 
the `fundamental' gapless modes of the symplectic symmetry class.

\paragraph{Weak coupling limit:}

Results for weak couplings, when the tunneling lengths are much larger than the phase-coherence lengths, are more interesting. They depend on both, the position of the 
Fermi level and system-size. This reflects the presence of additional small-gap modes
besides the fundamental soft modes of the symplectic class.
The precise number and gap-values of the former 
depend on the position of the Fermi level, and 
 gaps introduce new length scales, $L_\Delta$, 
which should be compared to the system-size.

 {\it $l_{\varphi S/B}\gg L_\Delta$:---} If this condition is fulfilled, contributions from small-gap modes can be neglected and  
\begin{align}
\alpha_S=1/2,\;\alpha_B=1/2, \qquad \beta_S=2,\;\beta_B=2.
\end{align} 
Physical properties are again 
determined by the fundamental soft modes of the symplectic class,  
which at weak coupling exist separately in bulk and surface. The relation $\beta=4$ also holds for the CFs in the universal regime ($l_{\varphi S/B}\gg L$) as long as the system size exceeds $L_\Delta$.

 {\it $l_{\varphi S/B}\ll L_\Delta$:---} In this limit, 
small gap-modes give contributions comparable
to those of the fundamental singlet modes, 
and have to be considered separately for small and large Fermi energies in the bulk.
In case of the former
\begin{align}
&\alpha_S=1/2, \alpha_B=-1, \;\;\beta_S=2,\;\beta_B=8, \quad \epsilon_{F,B}-M\ll M.
\end{align} 
The value $\beta=10$ signals the presence of each five soft or small gap modes in the
diffuson and Cooperon channels, respectively, while
$2\alpha=-1$ indicates that these five modes 
are composed of three spin-triplet modes in the bulk and two spin-singlet modes, one in the bulk and one on the surface. 
That is, at small Fermi energies three additional bulk spin-triplet modes contribute.
Notice that only the combined information 
on the interference correction to conductivity 
and conductance fluctuations provide a complete picture of the underlying physics. 
As from the interference corrections to conductivity 
alone one may, e.g., erroneously conclude on 
the presence of a single soft surface mode. In the universal regime, $\beta=10$ holds even under the somewhat weaker condition $L\ll (L_\Delta, l_{\varphi S/B})$.
In the limit of large Fermi energies, 
\begin{align}
\alpha_B=1,\; \alpha_S=1/2 \quad \beta_S=2,\;\beta_B=4, 
\quad \epsilon_{F,B}-M\gg M.
\end{align} 
Here, the value $\beta=6$ is understood from the presence 
 of each three soft or small gap modes in the diffuson and Cooperon channels, which  
now all manifest in the spin-singlet channel, i.e. leading to WAL $2\alpha=3$. Again, in the universal regime the condition for $\beta=6$ is somewhat weaker, $L\ll (L_\Delta, l_{\varphi S/B})$.

\paragraph{Intermediate coupling:}

The complete result for interference corrections to the Drude conductivity covering 
arbitrary couplings can be found in Sec.~\ref{sec:Interference_corrections}, 
Eq.~\eqref{eq:dg1main}
for small and large Fermi energies.
The corresponding results for the universal conductance fluctuations can be found in 
Sec.~\ref{sec:conductance_fluctuations}, Eq.~\eqref{eq:genresult}.

\subsubsection{TI thin film with two active surfaces}

\paragraph{Strong coupling limit:}
Suppose that the system consisting of the surface S1 and the effective bulk/S2 subsystem is in the strong coupling limit. Then, one finds
\begin{align}
\alpha=1/2,\quad \beta=2,
\end{align}
independent of the Fermi energy. As for the case with one active surface, these values of $\alpha$ and $\beta$ have their origin in the two fundamental soft modes of the symplectic symmetry class, the diffuson and the Cooperon.
\paragraph{Weak coupling limit:}
In the weak coupling limit, one obtains
\begin{align}
\alpha_S=1/2,\;\alpha_B=1/2,\quad  \beta_S=2,\; \beta_{B}=2,\quad  
\end{align}
independent of the position of the Fermi energy. The main difference to the corresponding case for one active surface is that the relation between the coherence lengths and $L_\Delta$ is not crucial here: The small-gap bulk modes are strongly suppressed due to the tunneling coupling to surface S2, only the fundamental modes in the bulk and in the surface S1 remain active.

\paragraph{Intermediate coupling:}
Eq.~\eqref{eq:genresult} of Sec.~\ref{sec:conductance_fluctuations} can be used to describe the intermediate coupling regime for the CFs. The relevant formula for the magneto-conductance is Eq.~\eqref{eq:gen2} of Sec.~\ref{sec:Interference_corrections}.

\subsubsection{Temperature dependence}
\label{sec:TD}

The results presented above are applicable if the thermal length $L_T=\sqrt{D/T}$ is the largest length scale in the problem. In this case, temperature broadening can be neglected and the CFs depend on temperature only through the coherence length(s). In practice, this condition may not always be fulfilled. The formalism can in principle be adapted to this situation.\cite{Altshuler86,Lee87,Akkermans2007} While we did not perform such a calculation explicitly in this work, we can draw certain conclusions about the expected dependence on $L_T$ from previous works on CFs in metallic systems. Previous results suggest the following generalization of our results: (i) If both $L_T$ and all relevant coherence lengths are much larger than the system size, then the results in the universal limit can be expected to remain valid. (ii) If the coherence lengths are the shortest length scales, then non-universal results are expected to carry over. (iii) If $L_T$ is the smallest relevant length scale, then $L_T$ essentially takes the role otherwise played by the coherence lengths in the non-universal limit. For example, the $(l_\varphi/L)^2$ dependences can be expected to be replaced by a $(L_T/L)^2$ dependence (it is known, however, that logarithmic corrections exist in 2D and that prefactors are modified). Nevertheless, the general scheme for determining the relevance of the different diffusion modes should remain intact.

\section{Soft modes}
\label{sec:soft_modes}
We proceed with a discussion of soft diffusion modes in the system. 
Readers not interested in technical details may proceed 
with Secs.~\ref{sec:Interference_corrections} and~\ref{sec:conductance_fluctuations}. We first review in Sec.~\ref{SMsurface} surface diffusion modes in the absence 
of bulk-surface coupling. This allows us to introduce the general formalism 
in a situation where the number of involved states is small.
We then proceed in Sec.~\ref{SMbulk} with a discussion of bulk diffusion-modes
in the different parameter regimes, and discuss 
in Sec.~\ref{SMcoupled}
consequences of a finite bulk-surface coupling.  
We closely follow Ref.~\onlinecite{Garate12}, who 
studied soft-modes of the Cooperon type relevant for WL/WAL 
corrections to the Drude conductivity, 
 and extend the analysis to diffuson soft-modes relevant for CF.

\subsection{Surface modes}
\label{SMsurface}

Taking into account disorder-scattering in the Born approximation, 
the single-particle propagator on the TI surface takes the form 
\begin{align}
\label{Gmatrix}
 G^{R/A}_{\epsilon,S}(\bold{k})
&=\left[\epsilon-\hbar v {\bf k}\boldsymbol{\sigma}\pm i\hbar /{2\tau_{0S}}\right]^{-1},
\end{align}
where the elastic scattering time   
\begin{align}
\label{surface:elastic-scattering-time}
   \frac{\hbar}{\tau_{0S}}  
    &=
        2\pi u^S_0  
        \int (d^2 k')
    \left| \left< \mathbf{k}  | \mathbf{k'} \right> \right|^2
    \delta( \epsilon_{F,S} - \epsilon_{\bold{k}'})		
    = 
     \pi u^S_0 \nu_S,
\end{align}
and for notational convenience we write $\bold{k}\equiv\bold{k}_\parallel$, 
the two-dimensional momentum of surface states. 
Here $\nu_S$ is the density of states at the Fermi-energy 
 per band and unit volume,
 $(d^dk)=d^d k/(2 \pi)^d$, and we already mentioned that 
 throughout the work we assume weak disorder $\epsilon_{F,S}\tau_{0S}\gg \hbar$. 
The single-particle propagator~\eqref{Gmatrix} 
decays on a scale set by the mean free 
path $ \ell_{0S}= v_S\tau_{0S}$, and the dynamics on longer length-scales is 
governed by soft modes in the system. 
The latter describe the combined scattering of particle- and hole-excitations 
from the same scattering centers and are reviewed next.
For later convenience we also recall
 the 
  transport scattering time 
  ($\hat{\bf k}={\bf k}/|{\bf k}|$) 
\begin{align}
\label{eq:scattering_time_surface}
\frac{\hbar}{\tau_S}&=2\pi u^S_0\int (d^2 k') 
(1-\hat{\bf k}\cdot\hat{\bf k'}) \left|\left< \mathbf{k}  
|\mathbf{k'} \right>\right|^2 \delta(\epsilon_{F,S}-\epsilon_{\bf k'}),
\end{align}
 and which  
relates as $\tau_S=2\tau_{0S}$.

\subsubsection{Surface-diffuson}

\begin{figure}[t]
\begin{center}
\includegraphics[scale=0.28]{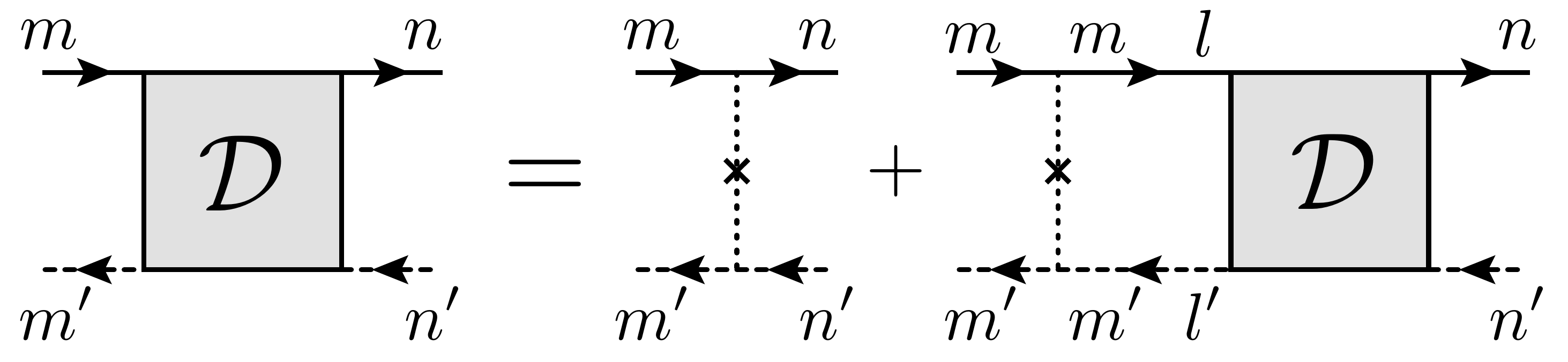}
\end{center}
\caption{
Diagrammatic representation of the Bethe-Salpeter equation 
for the surface-diffuson, Eq.~\eqref{surface:Bethe-Salpeter-original-basis}, 
expressed in the basis of spin-eigenstates. 
Solid (dashed) lines represent retarded (advanced) Green's functions, 
and crosses with dotted lines disorder scattering.~\cite{AGD63}}
\label{figure:BS_diffuson}
\end{figure}

Diffuson-modes $\mathcal{D}$ describe the combined  
propagation of a particle and a hole following the same 
sequence of scattering events, as  
encoded in the Bethe-Salpeter equation diagrammatically 
depicted in Fig.~\ref{figure:BS_diffuson}, 
\begin{align}
\label{surface:Bethe-Salpeter-original-basis}
 {\cal D}^{mn}_{m'n'} (\mathbf{q}) 
 &= u^S_0 \delta_{mn} \delta_{m'n'}
 + \sum_{l,l'=\uparrow,\downarrow} 
 U^{ml}_{m'l'} (\mathbf{q}) {\cal D}^{ln}_{l'n'} (\mathbf{q}).
\end{align}
Here 
$U^{ml}_{m'l'} (\mathbf{q}) 
 = u^S_0 \int (d^2 k) 
   G^R_{ml}(\mathbf{k}) G^A_{l'm'}(\mathbf{k} - \mathbf{q})$, 
and equations are written in the basis of spin eigenstates of 
$\sigma_z$, $|\uparrow\rangle$ and $|\downarrow\rangle$, 
diagonalizing disorder. 
For our purposes, it is sufficient to restrict to zero-energy 
Green's functions,  
$G^{R/A}({\bf k})\equiv G^{R/A}_{\epsilon_{F,S}}({\bf k})$, 
   \begin{align}
   \label{gfsurface}
   G^{R/A}({\bf k})=\frac{\epsilon_{F,S}
   \pm i\hbar/2\tau_{0S}+\hbar v_S {\bf k}\boldsymbol{\sigma}}
   {\left(\epsilon_{F,S}\pm i\hbar/2\tau_{0S}\right)^2-(\hbar v_S k)^2}.
   \end{align}
 The explicit form of the matrix $U$ then reads
\begin{align}
\label{surface:U-matrix}
  U^{ml}_{m'l'}(\mathbf{q}) 
  &= 
  a(\mathbf{q}) \, \delta_{ml} \delta_{m'l'}   + 
  b_i(\mathbf{q}) \, 	\delta_{ml} \sigma^i_{l'm'}   
  \nonumber\\
&\quad 
+ 
  c_i(\mathbf{q}) \, 	\sigma^i_{ml} \delta_{m'l'}   
+ d_{ij}(\mathbf{q}) \, 	\sigma^i_{ml} \sigma^j_{l'm'},  
\end{align}
where summation over repeated indices $i,j=1,2$ 
is implicit, 
and we introduced the following coefficient functions 
\begin{align}
\label{surface-U-matrix-coefficients-diffuson}
a(\mathbf{q})
&=
1/2 - \ell_{0S}^2 q^2/4,
\nonumber\\ 
b_i(\mathbf{q})
&=
c_i(\mathbf{q})
=
-i \ell_{0S} q_i/4,
\nonumber \\
d_{ii}(\mathbf{q}) 
&=  1/4 - \ell_{0S}^2 ( q^2 + 2q_i^2)/16,
\nonumber\\
d_{12}(\mathbf{q})
&=
d_{21}(\mathbf{q})
=
 - \ell_{0S}^2 q_xq_y/8,
\end{align}
We only kept leading orders in $q_i$.  
Notice that ${U}$ is not Hermitian 
and thus has a different set of left and right eigenvectors.
 
The formal solution to 
Eq.~\eqref{surface:Bethe-Salpeter-original-basis} reads
 \begin{align}
\label{surface:bethe-salpeter-cooperon-matrix}
	{\cal D}_\mathbf{q} 
	&= u^S_0 \left( \openone_{4} - {U}_\mathbf{q} \right)^{-1},
\end{align}
and we next identify most singular contributions in the limit $q\rightarrow 0$. 
To this end it is convenient to introduce the matrices of left and right eigenvectors of $U_{\bf q}$, 
denoted as $G_{\bf q}$ and $F_{\bf q}$, 
for which we fix normalization $G_{\bf q} F_{\bf q}=\openone_{4}$. 
This allows us to decompose the diffuson as 
$\mathcal{D}_{\bf q}=F_{\bf q}\hat{\mathcal{D}}_{\bf q} G_{\bf q}$,
where 
$\hat{\mathcal{D}}_{\bf q}^{-1}=G_{\bf q}(\openone_{4} -U_{\bf q})F_{\bf q}/u^S_0$ 
evaluates to
\begin{align}
\hat{\mathcal{D}}_{\bf q}^{-1}&=\frac{1}{u^S_0}\begin{pmatrix}
	    D_S q^2 \tau_{0S} & 0 					& 0 					& 0 \\
	    0 		 & \frac{1}{2} + \frac{D_S q^2 \tau_{0S}}{8} & 0 					& 0 \\
	    0 		 & 0					& \frac{1}{2} - \frac{D_S q^2 \tau_{0S}}{8}	& 0 \\
	    0 		 & 0					& 0					& 1
	\end{pmatrix},
	\label{Dsurface}
\end{align}
with $D_S=v_S^2\tau_S/2$.  
Eq.~\eqref{Dsurface} reveals the presence of one soft mode with vanishing 
eigenvalue in the long wave-length limit $q\rightarrow 0$. 
Disregarding massive modes 
we project onto the corresponding eigenspace 
and find ${\cal D}_\mathbf{q} 
        \simeq \frac{u^S_0}{D_S q^2 \tau_{0S}}
	F_0 
	{\rm diag}(1,0,0,0)
	G_0$. Neglecting then non-singular
momentum-dependencies of $F$ and $G$, we 
arrive at
\begin{align}
\label{surface:cooperon-matrix}
	{\cal D}_\mathbf{q} 
        \simeq 
	\frac{u^S_0}{2 D_S q^2 \tau_{0S}}
	\begin{pmatrix}
	    1		 & 0 		& 0 		& 1 \\
	    0 		 &  0		& 0		& 0 \\
	    0 		 & 0		&  0		& 0 \\
	    1 		 & 0		& 0		& 1
	\end{pmatrix},
\end{align}
which is diagonalized by the state $
  \frac{1}{\sqrt{2}} 
  \left(
    \left| \uparrow \uparrow \right> + \left| \downarrow \downarrow \right>
  \right)$. Despite its unusual basis-state, Eq.~\eqref{surface:cooperon-matrix} describes the spin-singlet singlet diffuson. 
  (Recall that the diffuson is formed of retarded and advanced propagators, see also Ref.~\onlinecite{Akkermans2007}).
  For further calculations it is also convenient to express the diffuson in the eigenbasis of the Dirac-Hamiltonian,
  \begin{align}
\label{surface:diffuson-transformation}
   {\cal D} (\mathbf{k}_1, \mathbf{k}_2, \mathbf{q}) 
   &= 
   \sum_{m,m'=\uparrow,\downarrow}\sum_{n,n'=\uparrow,\downarrow} 
   \left<   \mathbf{k}_1 | m  \right> \left< n|{\bf k}_2+{\bf q}\right>\times \nonumber\\
   &\times \left< m'|{\bf k}_1-{\bf q}  \right>\left<{\bf k}_2|n'\right>
   {\cal D}^{mn}_{m'n'} (\mathbf{q}),
\end{align}
and we recall that we assume $\epsilon_{F,S}>0$ and restrict to the upper band.
In the long-wavelength-limit we may neglect small momentum-transfer ${\bf q}$ 
in wave-function overlaps, and find 
\begin{align}
\label{surface:diffuson-matrix-element}
  {\cal D} (\mathbf{k}_1, \mathbf{k}_2, \mathbf{q})
  &= 
  \frac{\hbar}{2 \pi \nu_S \tau_{0S}^2}
  \frac{ 1}{D_S q^2}.
\end{align}
 Unlike for the calculation of average quantities, for the calculation of the CFs it becomes necessary to account for phase decoherence not only of the Cooperon but also of the diffuson. The corresponding phase relaxation rate will be introduced phenomenologically in this paper.

\subsubsection{Surface-Cooperon}

\begin{figure}[t]
	\centering
	\includegraphics[scale=0.28]{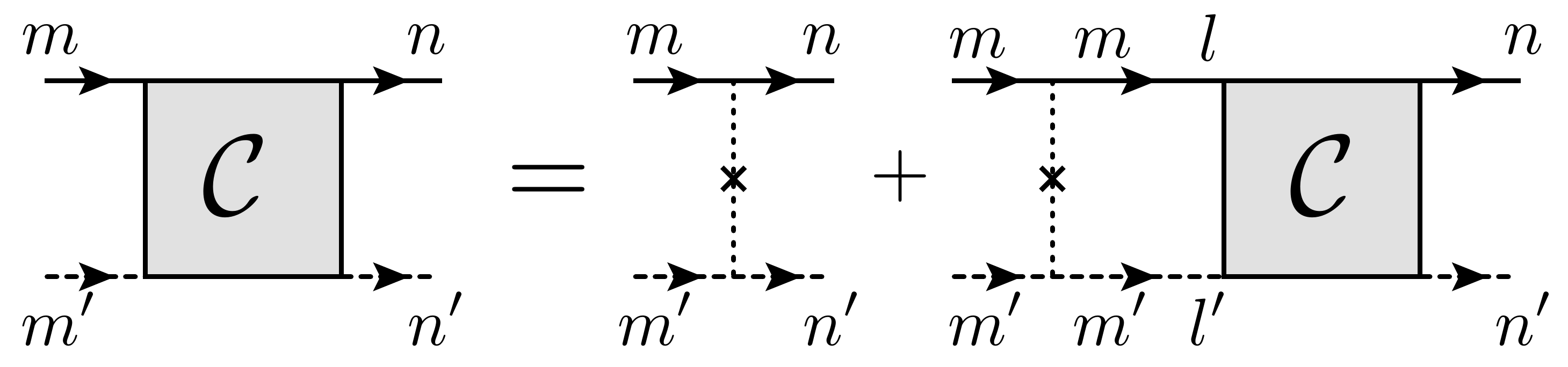}
	\caption{Bethe-Salpeter equation for the surface-Cooperon, 
	Eq.~\eqref{surface:Bethe-SalpeterC}, written in the basis of spin-eigenstates.}
	\label{figure:BS_Cooperon}
\end{figure}

Given time-reversal invariance of the system a second type of soft modes exists, which
 describes the collective propagation of a particle- and a hole-excitation 
along the same sequence of scattering events
in opposite direction. 
A diagrammatic representation of the Cooperon mode and 
its defining Bethe-Salpeter equation is shown in 
Fig.~\ref{figure:BS_Cooperon}. 
It takes the form
\begin{align}
\label{surface:Bethe-SalpeterC}
\mathcal{C}_{m'n'}^{mn}({\bf q})
&=
u^S_0\delta_{mn}\delta_{m'n'}+\sum_{l,l'}U^{ml}_{m'l'}({\bf q})\mathcal{C}^{ln}_{l'n'}({\bf q}),
\end{align}
where $U^{ml}_{m'l'}(\mathbf{q})
 = u^S_0 \int (d^2 k) 
   G^R_{ml} (\mathbf{k}) G^A_{m'l'}(-\mathbf{k} + \mathbf{q})$. 
   The matrix $U$ can be represented as, 
\begin{align}
\label{surface:U-matrix_for_C}
  U^{ml}_{m'l'}(\mathbf{q}) 
  &= 
  a(\mathbf{q}) \, \delta_{ml} \delta_{m'l'}   + 
  b_i(\mathbf{q}) \, 	\delta_{ml} \sigma^i_{m'l'}   
  \nonumber\\
&\quad 
+ 
  c_i(\mathbf{q}) \, 	\sigma^i_{ml} \delta_{m'l'}   
+ d_{ij}(\mathbf{q}) \, 	\sigma^i_{ml} \sigma^j_{m'l'},
\end{align} 
  with coefficients 
\begin{align}
\label{surface-U-matrix-coefficients-cooperon}
a(\mathbf{q})
&=
1/2 -  \ell^2_{0S} q^2/4,
\nonumber\\ 
b_i(\mathbf{q})
&=
-c_i(\mathbf{q})
=
i \ell_{0S} q_i/4,
\nonumber \\
d_{ii}(\mathbf{q}) 
&= -1/4 + \ell_{0S}^2 ( q^2 + 2q_i^2)/16,
\nonumber\\
d_{12}(\mathbf{q})
&=
d_{21}(\mathbf{q}) 
=
  \ell_{0S}^2 q_xq_y/8,
\end{align}
where we again only kept leading orders in $q_i$.

 A solution to Eq.~\eqref{surface:Bethe-SalpeterC} is derived
 following the previous discussion for the diffuson mode.\cite{Garate12}
The Cooperon is expressed in terms
of matrices $F_\mathbf{q}, G_\mathbf{q}$ of right- and left-eigenvectors 
of $U_{\bf q}$ (normalized by $G_{\bf q}F_{\bf q}=\openone_{4}$). 
That is, $\mathcal{C}_{\bf q}=F_{\bf q}\hat{\mathcal{C}}_{\bf q}G_{\bf q}$ 
where $\hat{\mathcal{C}}^{-1}_{\bf q}=G_{\bf q}(\openone_4-U_{\bf q})F_{\bf q}/u^S_0$, 
or explicitly,
\begin{align}
\hat{\mathcal{C}}^{-1}_{\bf q}=\frac{1}{u^S_0}
	\begin{pmatrix}
	    D_S q^2 \tau_{0S} & 0 					& 0 					& 0 \\
	    0 		 & \frac{1}{2} + \frac{D_S q^2 \tau_{0S}}{8} & 0 					& 0 \\
	    0 		 & 0					& \frac{1}{2} - \frac{D_S q^2 \tau_{0S}}{8}	& 0 \\
	    0 		 & 0					& 0					& 1
	\end{pmatrix}.
\end{align}
Concentrating again on the soft mode with vanishing eigenvalue as $q\rightarrow 0$,
we neglect non-singular momentum dependencies and 
 find $\mathcal{C}_{\bf q}\simeq \frac{u^S_0}{D{\bf q}^2 \tau_{0S}}F_0{\rm diag}(1,0,0,0)G_0$,
or
\begin{align}
\mathcal{C}_{\bf q}\simeq\frac{u^S_0}{2D_Sq^2\tau_{0S}}\begin{pmatrix}
	    0		 & 0 		& 0 		& 0 \\
	    0 		 &  1		& -1		& 0 \\
	    0 		 & -1		&  1		& 0 \\
	    0 		 & 0		& 0		& 0
	\end{pmatrix},
	\label{eq:cmatrixref}
\end{align}
which is diagonalized by the spin-singlet state 
$|0\rangle
=
  \frac{1}{\sqrt{2}} 
  \left(
    \left| \uparrow \downarrow \right> - \left| \downarrow \uparrow \right>
  \right)$. 
For latter convenience we give
the transformation to the spin-orbit basis in which disorder is diagonal, 
 \begin{align}
\label{surface:cooperon-transformation}
   {\cal C} (\mathbf{k}_1, \mathbf{k}_2, \mathbf{q}) 
   &= 
   \sum_{m,m'=\uparrow,\downarrow} 
      \sum_{n,n'=\uparrow,\downarrow}
   \left<   \mathbf{k}_1 		
   | m  \right>\left< n  | - \mathbf{k}_2 + \mathbf{q} 	\right>
  \nonumber\\
&\qquad \times   
\left< - \mathbf{k}_1 + \mathbf{q} 	| m' \right>
   \left< n' |   \mathbf{k}_2 			\right>
   {\cal C}^{mn}_{m'n'} (\mathbf{q}), 
\end{align}  
where we again projected onto the upper band. 
Neglecting non-singular ${\bf q}$-dependencies 
of the wave-function overlaps we arrive at the Cooperon in the spin-orbit basis, 
\begin{align}
\label{surface:cooperon-matrix-element}
  {\cal C} (\mathbf{k}_1, \mathbf{k}_2, \mathbf{q})
  &= 
  -\frac{\hbar}{2 \pi \nu_S \tau_{0S}^2}
  \frac{ e^{i (\phi_{k_1} - \phi_{k_2})}}{D_S q^2}.
\end{align}
As for the diffuson, sources of phase decoherence for the Cooperon mode will not be treated 
microscopically in this paper, and corresponding rates are introduced phenomenologically.

\subsection{Bulk modes}
\label{SMbulk}

The above discussion of soft surface-modes readily extends to the bulk. 
The single particle propagator in the self-consistent Born approximation 
reads
\begin{align}
G^{R/A}_{\epsilon}({\bf k})=\left(\epsilon - M \tau^z 
	- 
	\hbar v_B \mathbf{k} 
	\cdot 
	\boldsymbol\sigma \, \tau^x\pm {i\hbar}/{2\tau_{0B}}\right)^{-1},
	\label{eq:Green_bulk}
\end{align}
and the elastic and transport scattering times are~\cite{Garate12} 
\begin{align}
\label{bulk:scattering-time}
\hbar/\tau_{0B}
	 &=
           \pi u^B_0 \nu_B
	   \left( 1 + M^2/\epsilon_{F,B}^2 \right),\\
	   \hbar/\tau_B&=(2/3)\pi u^B_0 \nu_B\left(1+2M^2/\epsilon_{F,B}^2\right).
\end{align}
Here, $\nu_B$ is the density of states per band and per unit volume at the bulk 
Fermi energy $\epsilon_{F,B}$ and we assume weak disorder $(\epsilon_{F,B}-M)\tau_{0B}\gg \hbar$.
The Bethe-Salpeter equation for the diffuson in the basis of eigenstates 
$|m,m'\rangle$ of $\tau_z\otimes \sigma_z$, 
where 
$m,m' \in \left\{  T\uparrow, T\downarrow, B\uparrow, B\downarrow \right\}$, 
is
\begin{align}
\label{diffuson-3D-Bethe-Salpeter-so-basis}
	 D^{mn}_{m'n'} (\mathbf{q}) 
	 =
	  u^B_0
	  \delta_{mn} \delta_{m'n'}
	 +
	  \sum_{l,l'}
	  U^{ml}_{m'l'} (\mathbf{q}) 
	  D^{ln}_{l'n'} (\mathbf{q}), 
\end{align}
with
$U^{ml}_{m'l'} (\mathbf{q}) 
=u^B_0 \int (d^3 k)
G^{R}_{ml}(\mathbf{k})G^{A}_{l'm'}(\mathbf{k} - \bold{q})$, 
and
$G^{R/A}({\bf k})\equiv G^{R/A}_{\epsilon_F}({\bf k})$. $U$  
 is expanded as
\begin{align}
\label{3D-bulk:diffuson-U-matrix}
U^{ml}_{m'l'} 
	 &= 
	 a \, \delta_{ml} \delta_{m'l'}
	 + \sum_\mu b_\mu \delta_{ml} \Lambda^\mu_{l'm'}
\nonumber \\
	 &\qquad
	 + \sum_\mu c_\mu \Lambda^\mu_{ml} \delta_{m'l'} 
	 + \sum_{\mu\nu} d_{\mu\nu} \Lambda^\mu_{ml} \Lambda^\nu_{l'm'},
\end{align}
with $\Lambda^i=\sigma^i\tau^x$ for $i\in \{1,2,3\}$, 
$\Lambda^4=\tau^z$, and the summation in 
$\mu$, $\nu$ runs from $1$ to $4$. 
Explicit expressions for coefficient functions 
$a, b_\mu, c_\mu, d_{\mu\nu}$ are given 
in Appendix~\ref{app:bulk_diffuson}.

The formal solution to Eq.~\eqref{diffuson-3D-Bethe-Salpeter-so-basis}
 now results from the inversion 
$\mathcal{D} = u^B_0( \openone_{16} - \hat{U} )^{-1}$.
Diagonalization of the non-Hermitian $\hat{U}$, similar to Eq.~\eqref{Dsurface},
and expansion in 
 small momentum $\bold{q}$ reveals one gapless mode  
with eigenstate in the spin-singlet and orbital-triplet sectors, see Table~\ref{fsm}
and Appendix~\ref{app:bulk_diffuson} for details.   
This is the fundamental soft mode of the symplectic class, protected 
by particle-number conservation. (As already mentioned above, 
this protection does, however, 
not apply to diffusons describing ensemble fluctuations like CFs).

Additional modes with parametrically small masses emerge  
in the limit of large and small Fermi-energies. Specifically, 
in the limit $(\epsilon_{F,B} - M)/ M \ll 1$ there are three modes with masses 
$\Delta_{g1} = (2/9)(1 - M/\epsilon_{F,B})^2 \tau_{0B}^{-1}$. These are spanned by 
the eigenvectors given in Table~\ref{asm}.  
The mass $\Delta_{g1}$ is due to spin-flip transitions induced by the 
spin-orbit interaction of the bulk Hamiltonian. 
In the limit $(\epsilon_{F,B} - M)/ M \gg 1$ one finds one additional spin-singlet mode with small mass
$\Delta_{g2} = 2 (M/\epsilon_{F,B})^2 \tau_{0B}^{-1}$,
and an eigenvector that is in the triplet sector in orbital-space  
(see Table \ref{asm}). The mass $\Delta_{g2}$ 
is now due to inter-valley scattering 
$|T\rangle + | B \rangle \rightarrow |T \rangle - |B \rangle$, 
which, in turn, is due to the finite mass $M$. 
Finally, the diffuson in the eigenbasis of the clean bulk Hamiltonian
(upon projection onto the conduction band $\alpha,\beta=1,2$) 
takes the form, 
\begin{align}
	 &{D}^{\beta\beta'}_{\alpha'\alpha} (\mathbf{k}_1, \mathbf{k}_2, \mathbf{q}) 
	 = \sum_{m,m' } 
	 \sum_{n,n'} 
	   \left< \beta, 	\mathbf{k}_1 	| m \right >
	   \left< \alpha, 	\mathbf{k}_2	| n' \right>
\nonumber \\
	&\hspace{30 pt}
	   \times
	   \left< n  | \beta', 	\mathbf{k}_2 + \mathbf{q} \right>
	   \left< m' | \alpha', 	\mathbf{k}_1 -  \mathbf{q} \right>
	   {D}^{mn}_{m'n'} (\mathbf{q}).
\label{3D-bulk-diffuson-transformation}
\end{align}
The calculation of the Cooperon bulk mode proceeds along similar lines,\cite{Garate12} and 
is detailed in Appendix~\ref{app:bulk_cooperon}. 
The soft diffusion modes in bulk and surface are summarized in Tables~\ref{fsm} and \ref{asm}.

\begin{table}[t]
\centering
\begin{tabular}{l l l}
\hline\hline \noalign{\smallskip}
 {\bf Surface} \qquad
 & diffuson 
 &   
 $|{\cal D}_0\rangle 
\propto
(|\uparrow\uparrow\rangle + |\downarrow\downarrow\rangle)$
 \\
 \noalign{\smallskip}
\cline{2-3}\noalign{\smallskip}
& Cooperon \qquad
&
$|{\cal C}_0\rangle 
\propto
(|\uparrow\downarrow\rangle - |\downarrow\uparrow\rangle)$
 \\
 \noalign{\smallskip}
\cline{1-3}\noalign{\smallskip}
 {\bf Bulk} 
 & diffuson  
 &   
 $|D_0\rangle \propto
|\Lambda_0\rangle 
\otimes
(|\uparrow\uparrow\rangle + |\downarrow\downarrow\rangle)$
 \\
 \noalign{\smallskip}
\cline{2-3}\noalign{\smallskip}
& Cooperon 
&
$|C_0\rangle \propto
|\Lambda_0\rangle 
\otimes
(|\uparrow\downarrow\rangle - |\downarrow\uparrow\rangle)$\\
\noalign{\smallskip}\hline\hline
\end{tabular}
\caption{Eigenvectors of the fundamental soft modes. 
The existence of these modes is  
generally guaranteed by particle-number conservation, while 
in case of CFs they are also damped by a 
decoherence-rate  $1/\tau_\varphi$. 
  For the definition of the orbital vector $|\Lambda_0\rangle$ see Table~\ref{opwf}. 
}
\label{fsm}
\end{table}

\begin{table}[b!]
\centering
\begin{tabular}{l l l }
\hline\hline
\noalign{\smallskip}
${\bf \epsilon_{F,B}-M \ll M}\quad$ 
 & diffuson  
& 
$|D_{t,1}\rangle 
\propto
|\Lambda_1\rangle 
\otimes
|\uparrow\downarrow\rangle$
\\ && 
$|D_{t,0}\rangle 
\propto
|\Lambda_1\rangle 
\otimes
(|\uparrow\uparrow\rangle - |\downarrow\downarrow\rangle)$
\\
 && 
$|D_{t,-1}\rangle 
\propto
|\Lambda_1\rangle 
\otimes
|\downarrow\uparrow\rangle$
\\ \noalign{\smallskip}
\cline{2-3}\noalign{\smallskip}
& Cooperon 
&
$|C_{t,1}\rangle 
\propto
|\Lambda_1\rangle 
\otimes
|\uparrow\uparrow\rangle$
\\
&&
$|C_{t,0}\rangle 
\propto
|\Lambda_1\rangle
\otimes
(|\uparrow\downarrow\rangle + |\downarrow\uparrow\rangle)$
\\
&&
$|C_{t,-1}\rangle 
\propto
|\Lambda_1\rangle 
\otimes
|\downarrow\downarrow\rangle$
\\ \noalign{\smallskip}
\cline{2-3}\noalign{\smallskip}
& Mass 
&
$
\Delta_{g1} = 
( 1 - M/\epsilon_{F,B})^2 (2/9\tau_0)$
\\ \noalign{\smallskip}
\cline{1-3}\noalign{\smallskip}
${\bf \epsilon_{F,B}-M\gg M}$ 
 & diffuson  
 &   
$|D_s\rangle 
\propto|\Lambda_2\rangle 
\otimes
(|\uparrow\uparrow\rangle + |\downarrow\downarrow\rangle)$
 \\
 \noalign{\smallskip}
\cline{2-3}\noalign{\smallskip}
& Cooperon 
&
$|C_s\rangle \propto
|\Lambda_2\rangle 
\otimes
(|\uparrow\downarrow\rangle - |\downarrow\uparrow\rangle)$
\\ \noalign{\smallskip}
\cline{2-3}\noalign{\smallskip}
& Mass 
&
$
\Delta_{g2} = 
 (M/\epsilon_{F,B})^2 (2/\tau_0)$\\
 \noalign{\smallskip}
\hline\hline
 \end{tabular}
\caption{Eigenvectors of bulk soft-modes with parametrically small gaps. 
In the presence of decoherence these modes are damped by  
$\Delta_{gi}+1/\tau_{\varphi B}$.
For definitions of the orbital vectors 
$|\Lambda_1\rangle$ and $|\Lambda_2\rangle$, 
see Table~\ref{opwf}.
}
\label{asm}
\end{table}

\begin{table}[b!]
\centering
\begin{tabular}{l l}
\hline\hline\noalign{\smallskip}
$|\Lambda_0\rangle 
\propto
\alpha_+|TT\rangle
+
\alpha_-|BB\rangle$
&
$\alpha_\pm=(\epsilon_{F,B} \pm M)$
\\
\noalign{\smallskip}
\cline{1-2}\noalign{\smallskip}
$|\Lambda_1\rangle 
\propto \lambda_{1,T} |TT\rangle + \lambda_{1,B}  |BB\rangle\quad$
&
$\lambda_{1,T}=1 + {\cal O}([\epsilon_{F,B}/M - 1]^2)$
\\
&
$\lambda_{1,B} = {\cal O}(\epsilon_{F,B}/M - 1)$
\\
\noalign{\smallskip}
\cline{1-2}\noalign{\smallskip}
$|\Lambda_2 \rangle 
\propto
|TB\rangle
+
|BT\rangle$&
\\
\noalign{\smallskip}
\hline\hline
 \end{tabular}
 \caption{
 Orbital parts of the bulk soft-mode eigenfunctions.
}
\label{opwf}
\end{table}

\subsection{Bulk-surface coupling}
\label{SMcoupled}

With these preparations we now return to the gated films of interest, and discuss 
soft modes in the presence of a finite coupling between bulk and surface states.

Following the previous work Ref.~\onlinecite{Garate12}, we assume that coupling 
between surface and bulk states occurs randomly, i.e. at points $\{\mathbf{r}_i\}$ 
where the inhomogeneous depletion layer happens to be thin.
The local tunneling operator is modeled as a 
 random matrix within 
the symplectic symmetry class,
\begin{equation}
\label{coupling:tunneling-Hamiltonian}
	 \hat{T}
	 =
	\mathcal{N}\sum_{\{\mathbf{r}_i\}}
	 \sum_{\tau ={\rm T,B}}\sum_{l=0}^3
	  t_l^\tau(\bold{r}_i) \tilde{\sigma}^l \otimes 
	  	  \pi_\tau 
	    |{\bf r}_i^\parallel\rangle\langle{\bf r}_i|
	  +
	      {\rm h.c.}
\end{equation}
Here, $\tilde{\sigma}^l=(\openone_2,i\sigma_x,i\sigma_y,i\sigma_z)^t$ 
is a 4-vector composed of Pauli matrices acting on the spin degree of freedom, 
and operators $\pi_\tau$ project the orbital degree of freedom ($\tau={\rm T,B}$)
 onto the surface-state in the conduction band. 
 Tunneling-amplitudes $\{t_l^\tau(\bold{r}_i)\}_{l=0,...,3}$ are 
Gaussian distributed random variables, 
with vanishing mean and second moments 
$\langle t_l^\tau(\bold{r}_i) \, t_m^\sigma(\bold{r}_j) \rangle 
= 
\delta_{lm}\delta_{\sigma\tau} \delta_{ij} (t^2/2)$. 
The normalization factor is defined as $\mathcal{N}=\sqrt{VS/N_i}$, where 
$V$ and $S$ are the volume and the surface area of the film, respectively, and
$N_i=\sum_{\{{\bf r}_i\}}$ is the number of tunneling centers. 
The projection of ${\bf r}_i$ 
onto the two-dimensional surface is denoted as ${\bf r}_i^\parallel$.

The bulk-surface-coupling Eq.~\eqref{coupling:tunneling-Hamiltonian} opens an additional 
channel for elastic scattering, and thus modifies  
elastic scattering-rates of the single-particle propagators. 
The corresponding rates,
 accounting for the tunneling of bulk-states into the surface and vice versa,  
 are found from Fermi's golden rule  
\begin{align}
	 \hbar/\tau_{tB}
	 &=
	\frac{2 \pi}{V}	 
	 \int (d^2 p)
	 |\langle \alpha, \mathbf{k} |\hat{T} | \mathbf{p} \rangle |^2
	 \delta(\epsilon_{\alpha,k} - \epsilon_p),
	 \\
	 \hbar/\tau_{tS}
	 &=
	\frac{2 \pi}{S}	 
	\sum_{\alpha=1,2}
	 \int (d^3 k)
	 |\langle \mathbf{p} | \hat{T} | \alpha, \mathbf{k} \rangle |^2
	 \delta(\epsilon_{\alpha,k} - \epsilon_p),
\end{align}
and calculate to 
\begin{align}
\label{coupling:surface-to-bulk-tunneling-rate}
	 \hbar/\tau_{tB}
	 &=2 \pi S \nu_S t^2,
	 \qquad
	 \hbar/\tau_{tS}
	 = 4 \pi V \nu_B t^2. 
\end{align}
In the following we assume that scattering occurs predominantly due to 
disorder \emph{inside} the bulk and surface, so that the bulk and surface scattering rates 
are larger than the respective tunneling rates given 
in Eq.~\eqref{coupling:surface-to-bulk-tunneling-rate}. 
In appendix \ref{app:BSC}, we discuss the corrections to the single-particle propagators 
and soft modes induced by the bulk-surface coupling in detail, 
and here only summarize the results that are important for our further investigations. 

Bulk-surface coupling changes single particle propagators, 
Eqs.~\eqref{gfsurface} and~\eqref{eq:Green_bulk}, 
only in the form of a small correction to the elastic scattering rate, 
i.e.  $\tau_{0B/S}^{-1}\mapsto \tau^{-1}_{0B/S}+\tau^{-1}_{tB/S}$.
The impact on soft modes is more interesting, and it is instructive to recall their spin structure 
in absence of tunneling. 
Due to spin-orbit coupling two-particle propagators involving spin-triplet states are suppressed 
on the surface by spin-scattering processes, and 
surface soft-modes $\mathcal{D}_0$, $\mathcal{C}_0$ 
are spin-singlet modes. 
Similarly, the fundamental soft modes in the bulk, $C_0$ and $D_0$, are singlet.
Bulk modes with small gaps, 
on the other hand, can have finite spin in certain regimes. 
Indeed, a brief glance at Tables~\ref{fsm} and~\ref{asm} shows that 
at small and large  Fermi-energies the small-gap modes are 
spin-triplet, $C_{t,-1/0/1}$ and $D_{t,-1/0/1}$ 
and spin-singlet, $C_s$ and $D_s$, respectively. 
A finite coupling, Eq.~\eqref{coupling:tunneling-Hamiltonian}, 
then affects the soft modes in the following way: 
(i) the small-gap bulk modes acquire an additional damping 
$\Delta_{gi}+1/\tau_{\varphi B}\rightarrow \Delta_{gi}
+1/\tau_{\varphi B}+1/\tau_{tB}$ for $i\in\{1,2\}$ 
due to the tunneling to the surface, and (ii) the fundamental soft modes in bulk and surfaces hybridize. 
In the following, we elaborate on these two points.

(i)  Small-gap bulk soft modes do not couple to the soft surface modes. 
As explained in more detail in Appendix \ref{app:BSC}, 
even though in cases they manifest in the singlet channel 
the spin structure would allow for their coupling to 
surface soft modes, 
this coupling is not allowed by the orbital structure of the bulk modes 
(i.e. characterized by $|\Lambda_2\rangle$, see Table.~\ref{opwf}). 
This does not imply, however, their unaffectedness 
by a finite coupling, and they rather acquire an additional gap.
Specifically,   
at small Fermi energies $(\varepsilon_{F}-M)/M\ll 1$, 
they take the form
$X_t({\bf q})\equiv \langle X_t|C_{\bf q}|X_t\rangle$ 
with
\begin{align}
\label{eq:Ctx}
C_{t,1/0/-1}({\bf q})
&=
\frac{u_0^B/\tau_{0B}}{ D_B{\bf q}^2 + \Delta_{g1}+1/\tau_{\varphi B}+1/\tau_{tB}},
\\
D_{t,1/0/-1}({\bf q})
&=
\frac{u_0^B/\tau_{0B}}{D_B{\bf q}^2+\Delta_{g1}+1/\tau_{\varphi B}+{1}/\tau_{tB}},
\end{align}
where we introduced the phenomenological bulk-phase decoherence rate 
$1/\tau_{\varphi B}$.  At large Fermi energies $(\varepsilon_{F,B}-M)/M\gg 1$, on the other hand, 
slightly gapped bulk soft modes 
at finite coupling read
$X_s({\bf q})\equiv \langle X_s|C_{\bf q}|X_s\rangle$, with  
\begin{align}
C_s({\bf q})
&=
\frac{u_0^B/\tau_{0B}}{D_B{\bf q}^2+\Delta_{g2}+1/\tau_{\varphi B}+1/\tau_{tB}},
\label{eq:Cls}
\\
\label{eq:Ds}
D_s({\bf q})
&=\frac{u_0^B/\tau_{0B}}{D_B{\bf q}^2+\Delta_{g2}+1/ \tau_{\varphi B}+1/\tau_{tB}}.
\end{align}

(ii) At finite coupling 
 fundamental bulk- and surface-Cooperon modes,
$C^B_0({\bf q})=\langle C_0|C_{\bf q}|C_0\rangle$ 
and 
$C^S_0({\bf q})
\equiv 
\langle \mathcal{C}_0|\mathcal{C}_{\bf q}|\mathcal{C}_0\rangle$,
hybridize taking the form
(see Appendix \ref{app:BSC} for details)
\begin{align}
\label{eq:CX}
C^X_0({\bf q})
&=
\frac{u_0^X}{\tau_{0X} D_X}
\left[
\frac{A_X}{{\bf q}^2+q_{a}^2}+\frac{B_X}{{\bf q}^2+q_{b}^2}
\right],
\end{align}
where $X\in \{S,B\}$. 
Here, we introduced the momenta
\begin{align}
\label{eq:qaqb}
2q_{a/b}^2
&=
(l_S^{-2}+l_B^{-2})\pm\sqrt{(l_S^{-2}-l_B^{-2})^2+4l_{tS}^{-2}l_{tB}^{-2}},
\end{align}
with 
$D_Xl^{-2}_{X}=\left({1}/{\tau_{tX}}+{1}/{\tau_{\varphi X}}\right)$, 
$l^2_{tX}=D_{X}\tau_{tX}$, 
and coefficients
\begin{align}
A_{S/B}=1-B_{S/B}
&=
\frac{l_{B/S}^{-2}-q_{a}^2}{q_b^2-q_a^2}.
\end{align}
Similarly, fundamental bulk- and surface-diffuson modes,
$D_0^B({\bf q})\equiv \langle D_0|\mathcal{D}_{\bf q}|D_0\rangle$ 
and 
$\mathcal{D}_0^S({\bf q})
\equiv 
\langle \mathcal{D}_0|\mathcal{D}_{\bf q}|\mathcal{D}_0\rangle$, 
hybridize,
taking the same form as in Eq.~\eqref{eq:CX}.

Two limiting cases are particularly illuminating. 
If tunneling rates for both surface and  
bulk exceed respective phase decoherence rates, 
$\tau_{tS/B}\ll \tau_{\varphi S/B}$, 
then 
\begin{align}
\label{eq:strong1}
q_a^2\approx l_{tS}^{-2}+l_{tB}^{-2}, 
\end{align}
while $q_b$ is much smaller of the order of the inverse phase coherence lengths,
\begin{align}
\label{eq:strong2}
q_b^2
&\approx 
l_{\varphi S}^{-2}\frac{l_{tB}^{-2}}{l_{tB}^{-2}+l_{tS}^{-2}}
+
l_{\varphi B}^{-2}\frac{l_{tS}^{-2}}{l_{tB}^{-2}+l_{tS}^{-2}}.
\end{align} 
This will be referred to as the \emph{strong-coupling} case.
This implies that fundamental bulk and surface modes strongly hybridize 
and $q_b$ serves as a \emph{common} cut-off for the infrared divergence in ${\bf q}$. 
If, in turn, both phase relaxation rates exceed respective tunneling rates, 
$\tau_{tS/B}\gg \tau_{\varphi S/B}$, then we are dealing with the case of \emph{weak-coupling}, in which
\begin{align}
\label{eq:weak}
q_a
&\approx 
\max\left(1/l_{\varphi S},1/l_{\varphi B}\right),
\,\,
q_b\approx \min\left(1/l_{\varphi S},1/l_{\varphi B}\right),
\end{align}
and fundamental bulk and surface modes remain 
separate. Infrared divergences in ${\bf q}$ in this case are cut off by the respective bulk 
and surface inverse phase coherence lengths.

In view of points (i) and (ii), described above, the consequences of a finite coupling 
on the soft modes are  readily understood in the limits of weak and strong coupling-strengths. 
These limits 
correspond to cases in which scattering rates Eqs.~\eqref{coupling:surface-to-bulk-tunneling-rate} 
are smaller, respectively larger than the masses of the soft modes (\emph{including} phase decoherence rates) 
of the decoupled system,
and corresponding results were summarized in the previous section. 
That is,
only fundamental soft-modes survive strong bulk-surface coupling, and 
the gated film can be viewed as one single system with strong spin-orbit scattering 
with hybridized singlet modes $C_0^h$ and $D_0^h$, the fundamental soft modes
of the symplectic universality class.
At weak bulk-surface coupling, on the other hand, soft modes in surface and bulk remain 
(to a first approximation) decoupled.
At large bulk Fermi energies the gated film is then characterized by three diffuson and 
 three Cooperon soft-modes.  At small bulk Fermi energies, the system 
 has five diffuson and five Cooperon soft-modes, respectively. 
 At the lowest temperatures, when the bulk phase decoherence rate is much smaller than the gap 
 $\Delta_{g2}$ ($\Delta_{g1}$), the 
 ``fundamental" modes dominate the low-energy transport properties. 
 The above qualitative considerations are summarized in Table~\ref{tfsm} and intermediate regimes 
 can be inferred from the general expression derived above.

\begin{table}[]
\centering
\begin{tabular}{l  | l | l}
\hline\hline
\noalign{\smallskip}
 &\; {\bf diffuson} &\; {\bf Cooperon}
\\
\hline
{\bf strong coupling} & &
\\
\qquad $\epsilon_F-M\ll M\;$  &\; $D^h_0$ &\; $C^h_0$
\\
\qquad $\epsilon_F-M\gg M$  &\; $D_0^h$ &\; $C_0^h$
\\
\hline
{\bf weak coupling} & &
\\
\qquad $\epsilon_F-M\ll M$  &\; ${\cal D}_0,D_0,D_{t,1/0/-1}$ 
&\; ${\cal C}_0,C_0,C_{t,1/0/-1}$
\\

\qquad $\epsilon_F-M\gg M$  &\; ${\cal D}_0,D_0,D_s$\; 
&\; ${\cal C}_0,C_0,C_s$
\\
\noalign{\smallskip}
\hline\hline
  \end{tabular}
\caption{
Soft modes characterizing the gated films at different surface-bulk 
coupling strengths and positions of the bulk Fermi energy. 
Strong coupling refers to a situation where tunneling rates exceed phase decoherence rates, the weak coupling regime refers to the opposite limit. 
For strong coupling, fundamental bulk and surface diffusion modes 
hybridize and form modes denoted by $D_0^h$ and $C_0^h$, while at 
weak coupling they remain mostly separate. 
The evolution of fundamental soft modes in the full cross-over regime between 
strong and weak coupling is described by Eq.~\eqref{eq:CX} 
(and a corresponding formula for the diffuson). 
When expressed through the parameters $q_a^2$ and $q_b^2$, 
 strong and weak coupling regimes correspond to the limiting cases discussed 
around Eqs.~\eqref{eq:strong1}, Eqs.~\eqref{eq:strong2}, 
and Eq.~\eqref{eq:weak}, respectively.}
\label{tfsm}
\end{table}

\section{Interference corrections to transport}
\label{sec:Interference_corrections}

We have now prepared the stage to discuss interference corrections to the Drude conductivity and CFs. The former have been previously discussed in 
Ref.~\onlinecite{Garate12} and, some minor differences notwithstanding, we mostly confirm their results 
in this section.
CFs, to our knowledge, have not been addressed so far within a model taking into account finite coupling 
to bulk states, and will be discussed in the next section.  In this section, we will first be concerned with a single TI surface separated from 
the bulk by a depletion layer, the latter being due to the presence of a gate. Subsequently, we will complete the system and include the second TI surface as well.

Bulk and surface currents flowing in response to a spatially uniform electric field may be written as
${\bf j}_i=\sum_{j=B,S}\sigma_{ij}{\bf E}_j$, $i\in{B,S}$. For the sake of simplicity, 
we assume that both bulk and surface are subject to the same electric field 
${\bf E}={\bf E}_S={\bf E}_B$. 
In this set-up, the total current flowing through the sample is obtained as 
${\bf j}={\bf j}_B+{\bf j}_S=\sigma {\bf E}$ with $\sigma=\sum_{i,j=B,S}\sigma_{ij}$. 
Before we turn to the magneto-conductance of the gated film, 
it is instructive to briefly recall known expressions for the conductivities 
of bulk and surface considered as isolated systems. 

{\it TI bulk:}---The bulk conductivity reads 
\begin{align}
\sigma_{0B}
=
\frac{e^2\hbar}{2\pi}
\sum_{\alpha\beta}\int \frac{d^3k}{(2\pi)^3} 
v_{\alpha\beta}^x({\bf k})\tilde{v}^x_{\beta\alpha}({\bf k})G^R_\alpha({\bf k})G^A_\beta({\bf k}),
\end{align}
where 
${\bf v}_{\alpha\beta}({\bf k})=\langle \alpha {\bf k}|{\bf v}|\beta {\bf k}\rangle$ 
are matrix elements of the velocity operator fulfilling the relation 
${\bf v}_{\alpha\beta}({\bf k})=\hbar v^2\delta_{\alpha\beta}{\bf k}/{E_k}$. Matrix elements $\tilde{\bf v}_{\alpha\beta}({\bf k})={\bf v}_{\alpha\beta}({\bf k})(\tau_B/\tau_{0B})$ 
additionally include disorder-induced vertex corrections. 
With only the conduction band being active, 
the Drude result is obtained as $\sigma_{0B}=2e^2\nu_B D_B$. 
The result for the surface is found in an analogous way, $\sigma_{0S}=e^2\nu_S D_S$.

\begin{figure}[t]
	\centering
	\includegraphics[scale=0.33]{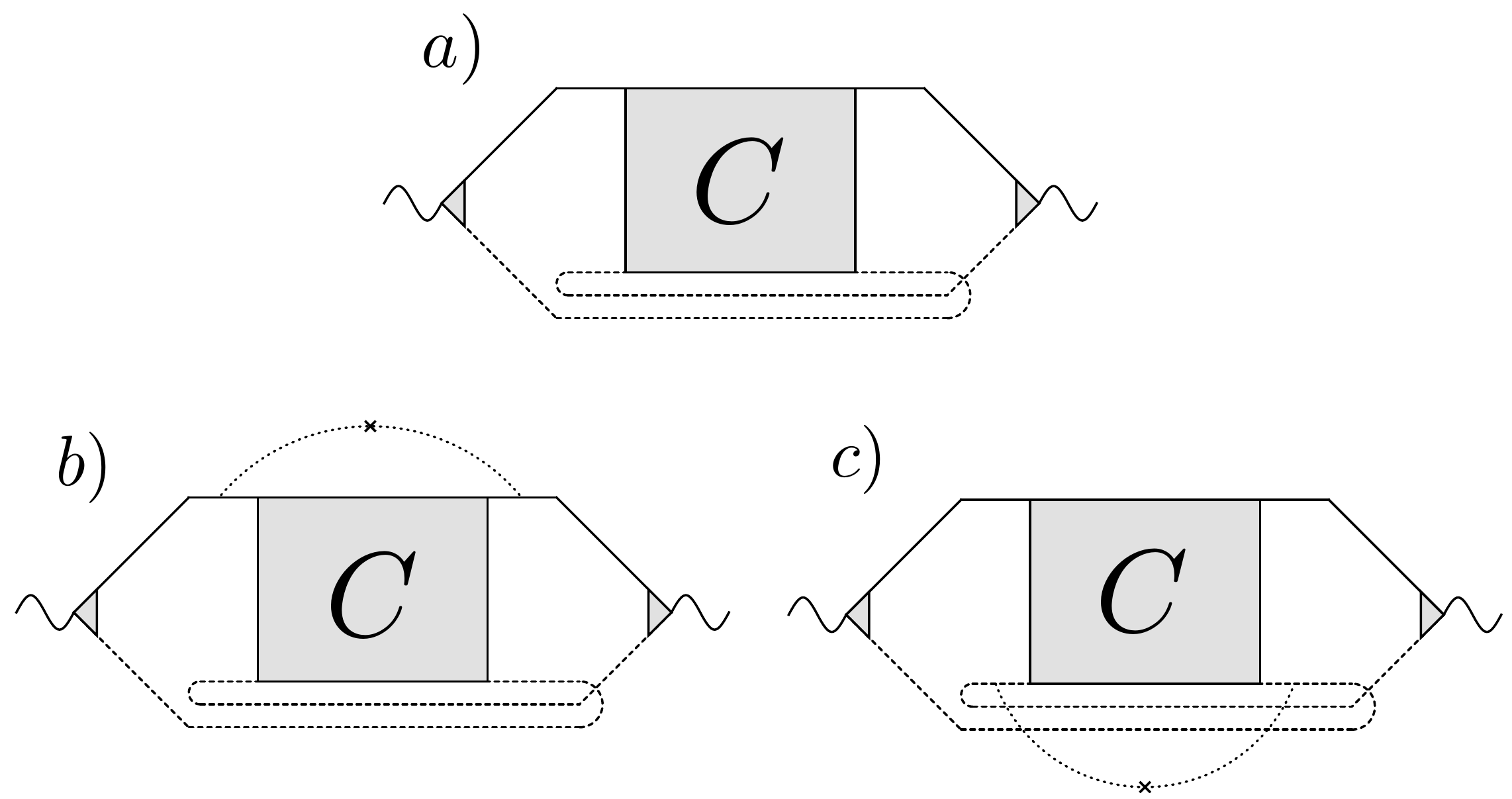}
	\caption{Diagrams giving the leading interference corrections to the Drude conductivity.}
	\label{figure:wl_corrections}
\end{figure}

Diagrams contributing to interference corrections to the Drude conductivity are 
depicted in Fig.~\ref{figure:wl_corrections}. 
For both bulk and surface, three diagrams need to be calculated,  
in the following accounted for by
 $\delta \sigma_{1X}$ (diagram a), 
and $\delta \sigma_{2X}$ (diagrams b and c)
with $X\in\{B,S\}$. Notice that the correction $\delta \sigma_{2X}$  becomes  
small in cases where the momentum-dependence of the disorder matrix 
elements is negligible.

Concentrating first on the bulk contribution, 
one finds~\cite{Garate12}
\begin{align}
\delta \sigma_{1B}
&=
-6\frac{e^2}{\hbar^2}\nu_B D_B\tau_B\tau_{0B}
\frac{1}{W}\int \frac{d^2 q}{(2\pi)^2}\overline{C}_B({\bf q}),
\label{eq:ds1}\\
\overline{C}_B({\bf q})
&=
\int (d{\bf n}_{\bf k}) 
({\bf n}^x_{\bf k})^2
\sum_{\alpha,\alpha'=1,2}C^{\alpha\alpha'}_{B,\alpha'\alpha}({\bf k}_F,{\bf k}_F,{\bf q}),
\end{align}
where $\int (d{\bf n}_{\bf k})$ denotes the normalized integral 
over the solid angle specifying directions of the $3d$-momenta on the Fermi surface,
and $W\ll l_{\varphi B}$ is the thickness of the film. 

For small Fermi energies, $(\varepsilon_{F,B}-M)/M\ll 1$, 
 three triplet Cooperon modes with gap $\Delta_{g1}$
contribute, besides the fundamental singlet mode, 
 to the averaged bulk Cooperon
\begin{align}
\overline{C}_B({\bf q})
=&
\frac{-\hbar}{6\pi \nu_B \tau_B^2}\left[\frac{1}{D_B{\bf q}^2+\tau^{-1}_{\varphi B}}
-
\frac{3}{D_B{\bf q}^2+\tau^{-1}_{\varphi B}+\Delta_{g1}}\right].
\label{eq:barC1}
\end{align}
Moreover, the momentum dependency of eigenstates $|\alpha {\bf k}\rangle$ is weak, and
 $\delta \sigma_{2B}$ is negligible compared to $\delta \sigma_{1B}$. 
 Performing then momentum integration in \eqref{eq:ds1} with $1/\sqrt{D_B\tau_H}$ as an upper cut-off, $\tau_H^{-1}=\tau_{\varphi B}^{-1}+2eD_BH/\hbar$, 
 one obtains in absence of an external magnetic field 
  $\delta G_B=W\delta \sigma_B$. 
In the presence of a weak (perpendicular) magnetic field
 $\Delta G_B(H)=G_B(H)-G_B(0)\approx \delta G_B(H)-\delta G_B(0)$, 
and the integration in ${\bf q}$ is replaced by an appropriate sum over 
Landau levels, resulting in 
\begin{align}
\Delta G_B(H)
&=
\frac{1}{2}G_q\left[f\left( x_{\varphi B} \right)-3f\left(x_1\right)\right],
\label{eq:DGsmall}
\end{align}
with 
$G_{q}=e^2/2\pi^2 \hbar$, $x_\alpha\equiv H_\alpha/H$ where
$H_{\varphi B}=\hbar/4eD_B\tau_{\varphi B}$,  
$H_1=(\hbar/4eD_B)\left[{1}/{\tau_{\varphi B}}+\Delta_{g1}\right]$, 
and 
the function $f(z)$ was defined in Sec.~\ref{sec:results}.
  The above expression simplifies in 
  two limiting cases,
\begin{align}
\label{dGDg}
\Delta G_B(H)
&={\alpha\over 2} G_q f\left(x_{\varphi B}\right), 
\end{align}
with
\begin{align}
\alpha
&=
\begin{cases}
1& \tau_{H}\gg \Delta_{g1}^{-1},
\\
-2& \tau_H \ll \Delta_{g1}^{-1}.
\end{cases}
\end{align} 
In the first case, 
the singlet mode gives the main contribution, 
while in the second triplet modes also need to be accounted for.

For large Fermi energies, $(\varepsilon_{F,B}-M)/M\gg 1$, 
the singlet mode with gap $\Delta_{g2}$ contributes besides
the fundamental Cooperon mode, and 
\begin{align}
\overline{C}_B({\bf q})
&=
\frac{-3\hbar}{8\pi\nu_B \tau_B^2}
\left[
\frac{1}{D_B{\bf q}^2+\tau^{-1}_{\varphi B}}
+
\frac{1}{D_B{\bf q}^2+\tau^{-1}_{\varphi B}+\Delta_{g2}}
\right].
\label{eq:barC2}
\end{align}
Moreover, in this limit, the momentum dependencies of the eigenstates 
are not negligible and $\delta \sigma_{2B}$ and $\delta \sigma_{1B}$ 
become comparable in size. 
 An explicit calculation gives~\cite{Garate12} 
 $\delta\sigma_{2B}=-\delta\sigma_{1B}/3$,   
 so that $\delta\sigma_B=2\delta\sigma_{1B}/3$. 
Quantum corrections are found in analogy to the 
case discussed above, and the magneto-conductance 
reads
\begin{align}
\Delta G(H)
&=
\frac{1}{2} 
G_{q}\left[f\left(x_{\varphi B}\right) + f\left(x_2\right)\right],
\label{eq:DGlarge}
\end{align}
where 
 $x_\alpha\equiv H_\alpha/H$ 
with
$H_2=(\hbar/4eD_B)\left[{1}/{\tau_{\varphi B}}+\Delta_{g2}\right]$, 
and the limits Eq.~\eqref{dGDg}
with
\begin{align}
\alpha
&=
\begin{cases}
1& \tau_H\gg \Delta_{g2}^{-1}, 
\\
2& \tau_H\ll \Delta_{g2}^{-1},
\end{cases}
\end{align}
where the first case is the result from the fundamental singlet mode, 
and the second accounts for the small-gap singlet mode.

{\it TI~surface:---}The contribution of a single surface reads~\cite{Garate12} 
\begin{align}
\delta\sigma_{1S}
&=
-4\frac{e^2}{\hbar^2}\nu_S\tau_S\tau_{0S}D_S
\int\frac{d^2q}{(2\pi)^2}\overline{C}_S({\bf q}),\\
\overline{C}_S({\bf q})
&=
\int(dn_{\bf k}) ({\bf n}_{\bf k}^x)^2 \mathcal{C}({\bf k}_F,{\bf k}_F,{\bf q}),
\end{align}
and $\overline{C}_S$ evaluates to
\begin{align}
\overline{C}_S({\bf q})
&=
-\frac{\hbar}{4\pi \nu_S\tau_{0S}^2}
\frac{1}{D_S{\bf q}^2+\tau^{-1}_{\varphi S}}.
\label{eq:barCS}
\end{align}
Noting that
$\delta \sigma_{2S}=-\delta\sigma_1/2$ 
and $\delta \sigma_S=\delta\sigma_{1S}/2$,  
the magneto-conductance is \cite{Lu11,Tkachov11,Garate12}
\begin{align}
\Delta G(H)
&=\frac{1}{2}G_qf\left(x_{\varphi S}\right),
\end{align}
where $x_{\varphi S}\equiv H_{\varphi S}/H$ 
with $H_{\varphi S}=\hbar/4eD_S\tau_{\varphi S}$.

\subsection{TI thin film with one active surface}

Results become more interesting once a finite tunneling between 
bulk and surface is introduced.
 Noting that~\cite{Garate12} the velocity vertex does not couple bulk and surface states 
 one anticipates that $\delta\sigma_{ij}=0$ for $i\ne j$. 
 That is, only Cooperon modes starting and ending or in the bulk or on the surface contribute.  
  These modes have been discussed in Sec.~\ref{SMcoupled}, and 
  to account for a finite tunneling we have to adjust the above calculation in the following way: 
(i) replacing contributions of the singlet Cooperon in $\overline{C}_{B(S)}$ 
in Eqs.~\eqref{eq:barC1},~\eqref{eq:barC2} (Eq.~\eqref{eq:barCS}) 
by the coupled mode $C_0^B$ ($C_0^S$) of Eq.~\eqref{eq:CX}, 
and (ii) inclusion of the tunneling-induced damping in the small-gap modes (see Eqs.~\eqref{eq:Ctx} and~\eqref{eq:Cls}). 
Employing then
 identities $A_B+B_B=A_S+B_S=1$ for coefficients in Eq.~\eqref{eq:CX}, 
 one finds the magneto-conductance of the coupled system
in the limits of small and large Fermi energies, 
\begin{align}
\Delta G(H)
&=
\frac{1}{2}G_q
\begin{cases}
f\left(x_a\right)+f\left(x_b\right)-3f\left(\tilde{x}_1\right), 
\\
f\left(x_a\right)+f\left(x_b\right)+f\left(\tilde{x}_2\right), 
\end{cases}
\label{eq:dg1main}
\end{align}
where the first and second line applies for $\varepsilon_F-M\ll M$ and 
$\varepsilon_F-M\gg M$, respectively.
Here $x_\alpha\equiv H_\alpha/H$, $\tilde{x}_\alpha=\tilde{H}_\alpha/H$ with
 $H_{a/b}=\hbar q_{a/b}^2/4e$ 
 and {$\tilde{H}_i=H_i+(\hbar/4eD_B\tau_{tB})$ 
 ($i\in{1,2}$), and 
 $H_1$ and $H_2$ 
 were defined below Eqs.~\eqref{eq:DGsmall} and~\eqref{eq:DGlarge}. 
 
It is instructive to consider the following limiting cases.  Both for 
 $\tilde{\tau}_H\gg \Delta_{g1}^{-1}$ at small Fermi energies and for 
 $\tilde{\tau}_H\gg \Delta_{g2}^{-1}$ at large Fermi energies, where $\tilde{\tau}^{-1}_H=\tau_{\varphi B}^{-1}+\tau_{tB}^{-1}+2eD_BH/\hbar$, one finds 
 \begin{align}
\Delta G(H)
&=
\frac{1}{2}G_q
\left[ 
f\left(x_a\right)+f\left(x_b\right)
\right].
\end{align}
Yet, in the opposite limit 
 $\tilde{\tau}_H\ll \Delta_{g1}^{-1}$ at small Fermi energies one gets
 \begin{align}
\Delta G(H)
&=
\frac{1}{2}G_q
\left[ 
f\left(x_a\right)+f\left(x_b\right) - 3f(\tilde{x}_{\varphi B})
\right],
\end{align}
where $\tilde{x}_{\varphi B}=(\hbar/3eD_B)(\tau_{\varphi B}^{-1}+\tau_{tB}^{-1})/H$,
and
for
 $\tilde{\tau}_H\ll \Delta_{g2}^{-1}$ at large Fermi energies one obtains
 \begin{align}
\Delta G(H)
&=
\frac{1}{2}G_q
\left[ 
f\left(x_a\right)+f\left(x_b\right) + f(\tilde{x}_{\varphi B})
\right].
\end{align}
The above results  are very similar to those obtained by Garate and Glazman. 
The difference occurs in the limit of large Fermi energies, 
where in our approach a coupling of small-gap singlet bulk Cooperons 
to the fundamental surface Cooperon mode 
is excluded, as explained in Sec.~\ref{SMcoupled}. 
This results in a change of the fields $H_a$ and $H_b$ in the case $\tilde{\tau}_B\gg \Delta_{g2}^{-1}$, compared to those found in Ref.~\onlinecite{Garate12}. 
In the corresponding fields entering the expression of these authors, 
the scattering rate $\tau_{tB}^{-1}$ enters with an additional factor of $2$ due to the additional decay channel. 
The limits of weak and strong coupling are further discussed in Sec.~\ref{sec:results}.

\subsection{TI thin film with two active surfaces}
\label{subsec:2active}
So far, we studied a gated TI film with only one active surface (S1). The key idea is that an external gate introduces a depletion layer between the bulk and the active surface, so that only a weak tunneling coupling remains. Yet, there is also the second surface (S2) which has not been treated explicitly so far. The previously derived results are still applicable if the second surface is passive. Indeed, considering first the system of bulk and S2 separately with the arguments given above, S2 becomes passive if (i) the coupling of bulk and S2 is weak (weak coupling limit) and (ii) the phasing decoherence rate in the bulk is much smaller than that in S2. Subsequently taking the first surface into account, one arrives at the results presented in the previous section.

In this section we discuss the case when surface S2 and the bulk are well connected and surface S2 is active. The main idea is again to consider S2 and the bulk as one subsystem and to subsequently couple it to S1. Due to the strong coupling in the S2/bulk subsystem, the fundamental hybridized singlet modes are dominant. When considering the coupling to S1, these modes acts analogously to a bulk singlet modes in the previous discussion, albeit with a modified decoherence rate. 
This modification reflects in the coefficients $q_{a}$ and $q_{b}$, which should now be calculated with the effective bulk/S2 hybrid coherence length. 
Notice that $q_b^2$ in Eq.~\eqref{eq:strong2}, evaluated for the bulk/S2 subsystem, acts as the effective $l_{\varphi B}^{-2}$ 
when calculating $q_a$ and $q_b$ for the S1/effective-bulk system. The final result reads
\begin{align}
\Delta G(H)=\frac{1}{2}G_q[f(x_a)+f(x_b)].
\label{eq:gen2}
\end{align}
As is clear from the result, only WAL is allowed in this case, since only the fundamental singlet modes survive the strong coupling to S2. A discussion of weak and strong coupling cases can be found in Sec.~\ref{sec:results}.

\section{Conductance fluctuations}
\label{sec:conductance_fluctuations}

We next turn to the discussion of the conductance fluctuations. We first briefly recall the calculation 
of conductance fluctuations in conventional weakly disordered metals in the presence of a single 
soft diffusion mode, and then turn to the system of interest, emphasizing differences to the former.

Consider spin-polarized electrons in a weakly disordered metal. For simplicity we assume 
isotropic scatterers for which the elastic free path and the transport mean path are identical, and the presence of 
a weak magnetic field that gaps out the Cooperon mode. 
Starting out from the Einstein relation, $\sigma \propto e^2 \nu D$, 
one notices that fluctuations of the  density of states  
and diffusion coefficient, $\delta \nu$ and $\delta D$, 
are typically uncorrelated. 
The conductance fluctuations then can be written as a sum of two independent 
contributions,  
$(\delta\sigma/\sigma)^2 = (\delta\nu/\nu)^2+(\delta D/D)^2$. 
This separation reflects in the diagrammatic representation, 
Fig.~\ref{figure:bulk_ucf_collective_diagram}, 
\begin{figure}[t]
	\centering
	\includegraphics[scale=0.36]{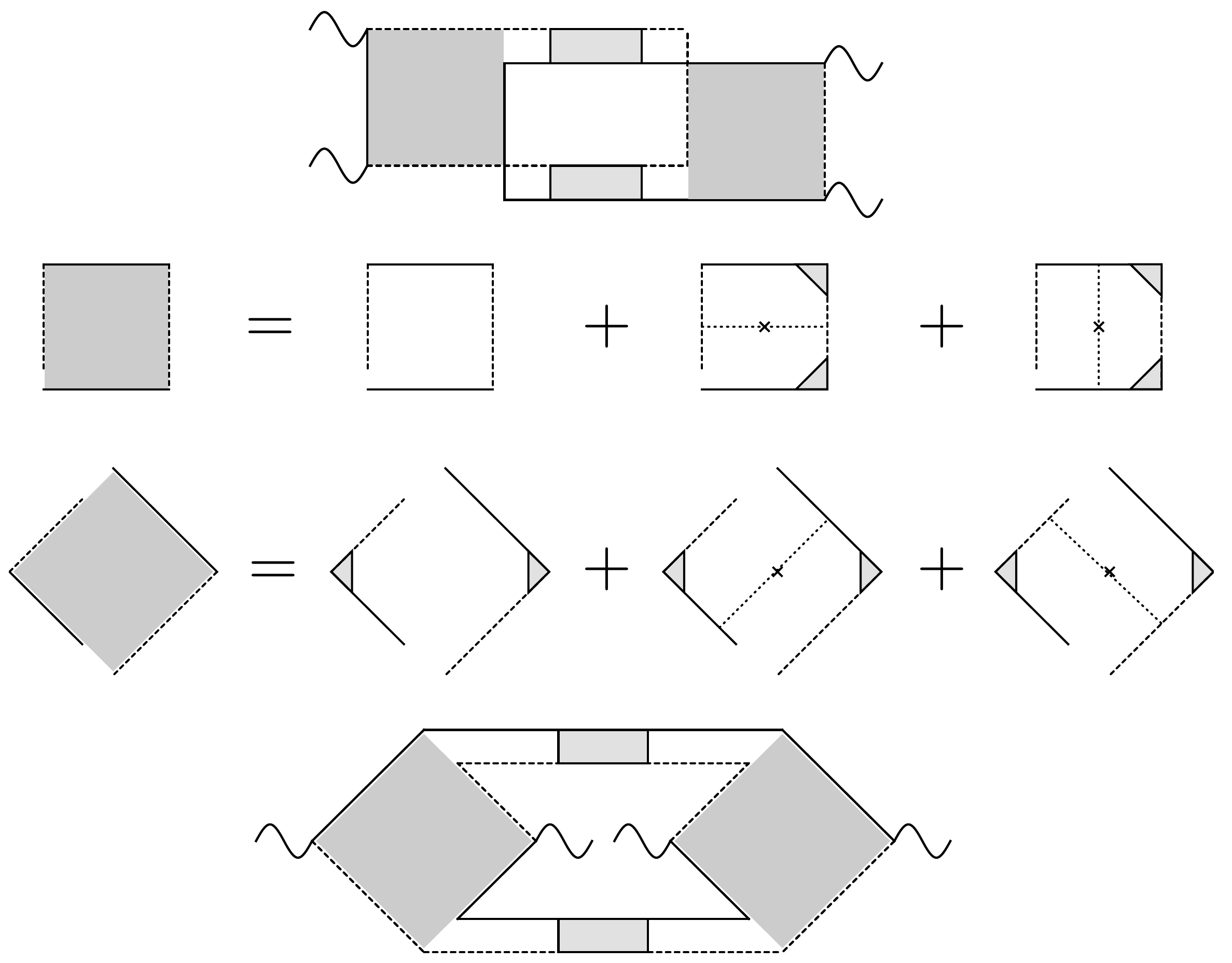}
	\caption{Diagrams contributing to the CFs. In the first and last line, 
	DOS-type and DCF-type diagrams, respectively, are shown. 
	The shaded regions correspond to Hikami boxes. 
	Full and dashed lines represent retarded and advanced Green functions, 
	respectively, dotted lines with crosses depict scattering on impurities, and 
	wavy lines with black triangles stand for current vertices with their 
	corresponding vertex corrections.}
	\label{figure:bulk_ucf_collective_diagram}
\end{figure}
consisting of two classes of diagrams that describe the respective fluctuations. 
The leading diagrams are made of two diffuson modes connected to two boxes containing the single Green's functions 
and a varying number of impurity lines. (Notice that the two mentioned classes of diagrams 
differ in the way current vertices are arranged inside the boxes.)
In case of the conventional weakly disordered metal, 
the contributions from soft modes and single-particle propagators 
decouple. Each diagram, therefore, factorizes into the product of a contribution 
from diffuson modes and from the respective box.
The latter can be summed independently for different diagrams entering the same class
(the `Hikami box') and determine the overall 
numerical prefactor of a given class of diagrams. 
The result of such calculation reads  
\begin{align}
\label{eq1a}
\delta G^2
&= 
(2+4) 
\left({e^2\over h}\right)^2 
\sum_{\bold{q}}
\mathcal{D}^2(\bold{q}),
\end{align}
where
$\mathcal{D}(\bold{q})
=
\left(L^2[\bold{q}^2+ L_\Delta^{-2}]\right)^{-1}$ is the zero-frequency diffuson in units of the
Thouless energy $E_{\rm th}=D/L^2$, 
and we have added a small mass 
$1/L_\Delta^2=\Delta/D$ to the propagator.
$e^2/h$ is the conductance quantum and the overall factor 
sums contributions from fluctuations of the density of states (contributing a factor 2)
and the diffusion constant (contributing a factor 4). The relative factor $1/2$ between the two classes of contributions 
is due to the renormalization of the Hikami-box by a single impurity line present for diagrams 
representing $\delta \nu$ 
but absent for those describing $\delta D$ in case of isotropic scatterers. 
The sum over momenta in Eq.~\eqref{eq1a} 
depends on the ratio of length-scales involved (we assume that the thermal length 
$\sqrt{D/T}$ is the largest length scale in the problem). 
Performing the sum over discrete momenta, one arrives at
\begin{align}
\label{eq1b}
\delta G^2 
&= 
 (e^2/h)^2 \, {\cal F}(L/L_\Delta),
\end{align}
where in the limit $L\ll L_\Delta$  
the function ${\cal F}(L/L_\Delta)$ 
(defined in Appendix \ref{subsec:final}, Eq.~\eqref{eq:Fsum}) 
becomes a universal number which  
 only depends on the sample geometry. In the opposite limit $L\gg L_\Delta$
it is quadratic in $L_\Delta/L$. 
Specifically, for the quasi two-dimensional sample with open boundary conditions 
 \begin{align}
 {\cal F}(L/L_\Delta)
 &\simeq
 \begin{cases}
0.093, \quad &L\ll L_\Delta,
\\
3/(2\pi)(L_\Delta/L)^2, \quad & L\gg L_\Delta.
 \end{cases}
 \end{align}
Eqs.~\eqref{eq1a} and~\eqref{eq1b} are the result for a single soft diffusion mode.
In the absence of a magnetic field the Cooperon mode gives an identical contribution, resulting 
in an additional overall factor $2$. 
Accounting further for the electron-spin, another overall
factor $2^2=4$ arises (in absence of spin-orbit scatterers) 
in Eqs.~\eqref{eq1a} and~\eqref{eq1b}. 
Finally, for anisotropic scattering, soft modes  
depend on the transport mean path which differs from 
elastic scattering length. The latter, on the other hand,
enters the Hikami boxes and,  
accounting for the renormalization of current vertices 
by diffuson modes, e.g. shown in Fig.~\ref{figure:bulk_ucf_collective_diagram}, 
the universality of CF is recovered in the limit $L\ll L_\Delta$.

We now return to the system of interest, 
first concentrating on the contribution from the bulk topological insulator. The more involved internal structure of the Hamiltonian Eq.~\eqref{intro_bulk_Hamiltonian} compared to the conventional 
weakly disordered metal is reflected in the band- and momentum-dependencies of the disorder matrix elements. 
The latter implies that, similar to the case of anisotropic scattering discussed above,
 vertex corrections to the current operators have to be accounted for. More crucially, however, it implies that
a factorization of contributions from diffusion modes and single particle propagators does no longer hold.
That is, instead of accounting for all diagrams contributing to one class of fluctuations 
in terms of a single Hikami box, one now has to calculate each diagram separately. 
There are nine non-vanishing diagrams contributing each to fluctuations of the 
density of states and the diffusion constant. 
In both cases these consist of one `bare' diagram in which single impurity lines are absent, 
and each four diagrams are dressed by a single or two impurity lines, see also 
Fig.~\ref{figure:bulk_ucf_collective_diagram}. Similar considerations apply to the 
CFs of an isolated topological-insulator surface 
(in this case we reproduce the results of Ref.~\onlinecite{Adroguer12}, see below).

\subsection{TI thin film with one active surface} 
For a gated film with finite bulk-surface coupling, we are interested in the fluctuations 
of the total conductance $G=G_S+G_B$. 
The calculation of the fluctuation contributions for the coupled system involves 
cross-correlations of the bulk and surface conductances, 
in addition to the individual fluctuations of the bulk and surface conductances. 
These can again be classified into fluctuations of the density of states and fluctuations 
of the diffusion coefficients. 
A more detailed explanation of the calculation is provided in appendix \ref{app:UCF}. 
Table~\ref{table:UCF} summarizes how each of the $2\times 9$ diagrams 
contributes to the fluctuation of the surface and bulk conductances as well as the 
cross-correlations for large and small Fermi energies. 
Here, we only summarize our findings,
\begin{align}
\label{eq2a}
\delta G^2
&= 
(2+4) 
\left({e^2\over h}\right)^2 
\sum_\alpha \sum_{\bold{q}}
X_\alpha^2(\bold{q}),
\end{align}
where $2+4$ results again from fluctuations of the 
diffusion constant and density of states, respectively.  
The sum is over all (zero-frequency) soft modes  (normalized by $E_{\rm th}$)
$X_\alpha$
\begin{align}
X_\alpha({\bfq})=\frac{1}{L^2}\frac{1}{{\bf q}^2+1/{L^2_{\Delta_\alpha}}}.
\end{align}
In Eq.~\eqref{eq2a}, 
$\alpha$ runs over all soft diffusion modes in the relevant regime, 
including the hybridized fundamental modes and soft spin-singlet and triplet modes, 
where applicable.

 Notice that the result \eqref{eq2a} is common to the limits of large ($\eps_{F,B}-M\gg M$) 
 and small ($\eps_{F,B}-M\ll M$) Fermi energies. 
 The relevant soft modes, however, differ in number and spin structure in these two limits 
 and also depend on the strength of the tunneling, as summarized in Table~\ref{tfsm}. 
 The fundamental singlet modes are active irrespective of the limit under consideration. 
Choosing the Cooperon as an example, the fundamental singlet mode(s) enter in the form 
\begin{align}
\sum_{\alpha=a,b}X^2_{\alpha}(\bfq)=\frac{1}{L^4}\left[\frac{1}{({\bf q}^2+q_a^2)^2}+\frac{1}{({\bf q}^2+q_b^2)^2}\right],
\end{align}
where $q_a$ and $q_b$ are defined in Eq.~\eqref{eq:qaqb} 
(The fundamental diffuson enters in an analogous form). 
We recall from Sec.~\ref{SMcoupled} that in the case of weak coupling the two terms 
labelled by $a$ and $b$ represent the separate fundamental bulk and surface soft modes, 
while in the case of strong coupling these modes hybridize with $q_b$ remaining small 
of the order of the inverse phase coherence length, while $q_a$ becomes large 
(of the order of the inverse tunneling length). We also recall that for the remaining 
(non-fundamental) soft modes $X_\alpha$ in Eq.~\eqref{eq2a} the coupling increases 
the mass and finally suppresses their contributions in the strong-coupling limit, 
where the hybridized 
fundamental modes remain the only active soft modes. 
In Appendix \ref{app:UCF}, Eq.\eqref{eq2a} is cast in a more explicit form for the interested reader.
Finally, we notice that, as is clear from the above discussion, a similar equation to Eq.~\eqref{eq2a} also holds for the isolated surface 
of the topological insulator, as well as for the isolated bulk, only
in this case the sum is over the corresponding surface soft modes or bulk soft modes (Appendix~\ref{app:UCF} provides more details for these cases as well).

We now collect pieces and discuss the conductance fluctuations in different limits of interest.
These are the limits of weak and strong bulk-surface coupling, and large and small Fermi energies, respectively.
Assuming again sufficiently low temperatures and that $\sqrt{D/T}$  is the largest length scale,
we start out from the general expression
\begin{align}
\delta G^2 
&=  
 (e^2/h)^2 \sum_{\alpha} {\cal F}(L/L_{\Delta_\alpha}),
 \label{eq:genresult}
\end{align}
and recall our discussion of soft modes in the different regimes. While formula \eqref{eq:genresult} 
is quite general and covers the entire range between the universal and non-universal regime, 
it is instructive to discuss these limiting cases, where simple results may be obtained.

\subsubsection{Universal regimes}

At sufficiently small temperatures the phase coherence length sets the largest length scale
and conductance fluctuations become universal. In the TI films we can identify universal 
regimes corresponding to the strong and weak coupling limits, respectively.
In both regimes 
we only need to count the number of soft modes that are present.  
 In all expression given below we assume that all coherence lengths exceed the system size.

{\it Strong coupling limit:---}A brief glance at Tables~\ref{fsm} and~\ref{asm} 
shows that for strong coupling
\begin{align}
\delta G^2 
&=  
 (e^2/h)^2  \times 0.093\times 2,
\end{align}
independent of the Fermi energy. Indeed, if the tunneling lengths are much smaller than the system size only the two fundamental hybridized Cooperon and diffuson modes give a sizable contribution.

{\it Weak coupling limit:---}At weak coupling and system sizes $L\ll L_{\Delta_\alpha}$ for all $\alpha$
\begin{align}
\delta G^2 
&=  
 (e^2/h)^2  \times 0.093 
 \times
 \begin{cases}
 10, & \epsilon_{F,B}-M\ll M
 \\
 6, & \epsilon_{F,B}-M\gg M.
 \end{cases}
\end{align} 
Here, all bulk and surface soft modes contribute.  In this limit, the surface is entirely decoupled from the bulk. 
 
If $L_{\Delta_\alpha}\ll L$ holds for all non-fundamental soft modes, then 
\begin{align}
\delta G^2 
&=  
 (e^2/h)^2   \times 0.093 
 \times
 4, 
 \label{eq:s2}
\end{align}
independent of the Fermi energy. 
This is the contribution of the two fundamental surface and the two fundamental bulk soft modes.

\subsubsection{Non-universal regime}

Finally, we consider the case of strong decoherence in which the 
 phase coherence length is much smaller than the system size and results for the conductance fluctuations become non-universal. 

In this case, one obtains 
\begin{align}
\delta G^2 
&=  
2\times\left(\frac{e^2}{h}\right)^2 \times  {3\over 2\pi}\left({L_\varphi\over L}\right)^2 
\label{eq:g2NU}
\end{align}
in the \emph{strong coupling} limit. Here, $L_\varphi$ is the effective phase coherence length obtained from Eq.~\eqref{eq:strong2} and the contribution comes from the two fundamental hybridized diffusion modes.

At \emph{weak coupling}, if the bulk and surface  phase coherence lengths $l_{\varphi B}$, $l_{\varphi S}$ are the shortest length scales in the problem, one finds
\begin{align}
\delta G^2 
&=  
\left(\frac{e^2}{h}\right)^2 \times  {3\over 2\pi}\left[\beta_B\times \left({l_{\varphi B}\over L}\right)^2+\beta_S\times \left({l_{\varphi S}\over L}\right)^2\right] 
\label{eq:gsqwc} 
\end{align}
where
\begin{align}
&\beta_B=8,\quad \beta_S=2,\quad  \epsilon_{F,B}-M\ll M,\\
&\beta_B=4,\quad \beta_S=2,\quad \epsilon_{F,B}-M\gg M.
\end{align}
In this limit, all soft modes give a similar contribution. 
If, however, the non-fundamental bulk soft modes are gapped out (the related lengths $L_{\Delta_\alpha}$ are much smaller than the phase coherence lengths) in the weak coupling limit, then $\beta_B=2$ in Eq.~\eqref{eq:gsqwc} for both large and small Fermi energies.

\subsection{TI thin film with two active surfaces}

The results presented in the previous section remain valid if the second surface (S2) is weakly coupled and passive, as explained in Sec.~\ref{subsec:2active}. If S2 is active and strongly coupled, however, all but the fundamental soft modes in the bulk/S2 subsystem are suppressed. This is the regime we will discuss in this section. Simple results can be obtained in the following limits.

\subsubsection{Universal regimes}
Here we discuss cases in which all phase coherence lengths exceed the system size $L$.

{\it Strong coupling limit:---} If $L$ is much larger than the tunneling lengths, then the combination of surface S1 and the effective bulk/S2 subsystem is in the strong coupling limit. In this case one finds
\begin{align}
\delta G^2 
&=  
 (e^2/h)^2   \times 0.093 
 \times
 2. 
\end{align}
This contribution originates from the hybridized Cooperon and diffuson modes of the film.

{\it Weak coupling limit:---}
This limit is reached if tunneling events between the surface S1 and the bulk/S2 subsystem are so rare that the tunneling length exceeds even the phase coherence length. It means that S1 and the bulk are well separated. Then only 4 modes remain active, two on the surface S1 and two in the bulk/S2 subsystem. Therefore, 
\begin{align}
\delta G^2 
&=  
 (e^2/h)^2   \times 0.093 
 \times
 4. 
\end{align}
This case is closely related to Eq.~\eqref{eq:s2}, only in the present situation the non-fundamental modes are inactive not due to their gaps but due to the strong tunneling to S2.

\subsubsection{Non-universal regime}
Here we consider cases in which the system size $L$ is by far the largest relevant length scale.

{\it Strong coupling limit:---} If all phase coherence lengths exceed the tunneling lengths, then one finds
\begin{align}
\delta G^2&=2\times \left(\frac{e^2}{h}\right)^2\times \frac{3}{2\pi} \left(\frac{L_{\varphi}}{L}\right)^2
\end{align}
where $L_\varphi$ is the effective phase coherence length resulting from the combination of the effective length for the B/S2 subsystem and the phase coherence length of the surface S1 weighted with the tunneling rates according to Eq.\eqref{eq:strong2}. This case is akin to Eq.~\eqref{eq:g2NU}, albeit with a new phase coherence length.

{\it Weak coupling limit:---} If the phase coherence length of the surface S1, $l_{\varphi S1}$, and the effective phase coherence length of the bulk/S2 subsystem, $l_{\varphi B/S2}$, are shorter than the tunneling length between bulk and surface S1, then
\begin{align}
\delta G^2 
&=  
\left(\frac{e^2}{h}\right)^2 \times  {3\over 2\pi}\left[2\times \left({l_{\varphi B/S2}\over L}\right)^2+2\times \left({l_{\varphi S}\over L}\right)^2\right],  
\end{align}
for both large and small Fermi energies. The difference to \eqref{eq:gsqwc} lies in the replacement $\beta_B\rightarrow \beta_{B/S2}=2$, which occurs due to the strong coupling to S2 irrespective of the position of the Fermi level.

\section{Conclusion}
\label{sec:conclusion}
In this work, we presented a combined study of magnetoresistance and CFs in thin gated TI films with bulk-surface coupling. For these quantities, the interpretation of experimental data is often complicated by their dependence on various characteristic length scales and energy scales of the system. Combined studies of both effects promise additional insight. The underlying reason is the sensitivity of both magnetoresistance and CFs at low temperatures to the number, spin-structure and masses of the slow diffusion modes in the system. At the same time, the interpretation of experimental results also requires theoretical predictions in different parameter regimes. In line with this reasoning, the main aim of this paper was to derive results for both effects based on the same theoretical model. To this end, we adopted the model proposed in Ref.~\onlinecite{Garate12}, where WL/WAL corrections to magnetoresistance were already calculated, and generalized the study to include CFs. 

The results for the magnetoresistance and the CFs can be characterized by the coefficients $\alpha$ and $\beta$ introduced in Eqs.~\eqref{eq:ab1} and \eqref{eq:ab2}. We studied TI insulator films with one or two active surfaces as detailed in Sec.~\ref{subsec:model}. For the model with two active surfaces we restricted our considerations to the case that (at least) one surface is strongly tunneling-coupled to the bulk. Generally speaking, the position of the bulk Fermi energy relative to the bulk gap determines the number of active diffusion modes in the bulk and has therefore a strong influence on the low temperature transport properties of these TI films. In this paper we studied two cases, large Fermi energies far exceeding the bulk band gap and small bulk Fermi energies that are close to the bottom of the conduction band. Several length scales are of crucial importance, the phase coherence lengths and tunneling lengths in bulk and surface(s) as well as additional length scales $L_{\Delta_\alpha}=\sqrt{D/\Delta_\alpha}$ related to the gaps $\Delta_\alpha$ of the (non-fundamental) bulk diffusion modes. 

Simple analytical results have been obtained in the strong coupling limit, when all phase coherence lengths are much larger than the tunneling lengths, or in the opposite case, the weak coupling limit. 
In the strong coupling limit, only the hybridized fundamental diffusion mode for the combined system of bulk and surface(s) is active and the result is $\alpha=1/2$ and $\beta=2$ both for one and two active surfaces. For our model of films with two active surfaces small-gap bulk modes are suppressed even in the weak coupling limit due to the frequent tunneling events between bulk and the surface that is always assumed to be strongly coupled. Correspondingly, the subsystem consisting of the bulk and the strongly coupled surface on one hand and the weakly coupled surface on the other hand contribute additively, leading to $\alpha_S=1/2$, $\alpha_B=1/2$, $\beta_S=2$, and $\beta_B=2$. A richer behavior is found for TI thin films with one active surface in the weak coupling limit as illustrated in Fig.~\ref{fig:wc}. One obtains different results depending on the position of the bulk Fermi energy and the relation between the dephasing lengths and the gap-related length scales. This regime has also been addressed in earlier works.
The values $\alpha_B$ in the weak coupling limit are in agreement   
with the study of Ref.~\onlinecite{Ostrovsky12} considering only the bulk Hamiltonian Eq.~(2), and 
coefficient $\beta_S$, describing conductance fluctuations for the surface, has previously been found 
in Ref.~\onlinecite{Adroguer12} where an isolated TI surface was studied. The results for weak localization coefficients $\alpha_S$ and $\alpha_B$ 
 for the surface and bulk coincide with those obtained in Ref.~\onlinecite{Garate12} 
 for the same model, and we here add   
 coefficients $\beta_S$ and $\beta_B$ 
to the picture.

A more detailed discussion of the main theoretical predictions obtained in this paper can be found Sec.~\ref{sec:results}, where we also comment on the temperature dependence. Expressions for the intermediate coupling regime have also been derived in this paper and can be found in the respective sections on interference corrections, Sec.~\ref{sec:Interference_corrections}, and universal conductance fluctuations, Sec.~\ref{sec:conductance_fluctuations}. Let us only remark here that the obtained predictions indeed confirm the potential of simultaneous measurements of magnetoresistance and CFs for resolving ambiguities in the interpretation of experimental results. We hope that in this way the present work can contribute to a better understanding of the low-temperature transport properties of TI materials.

\begin{figure}[tb]
	\centering
	\includegraphics[scale=0.53]{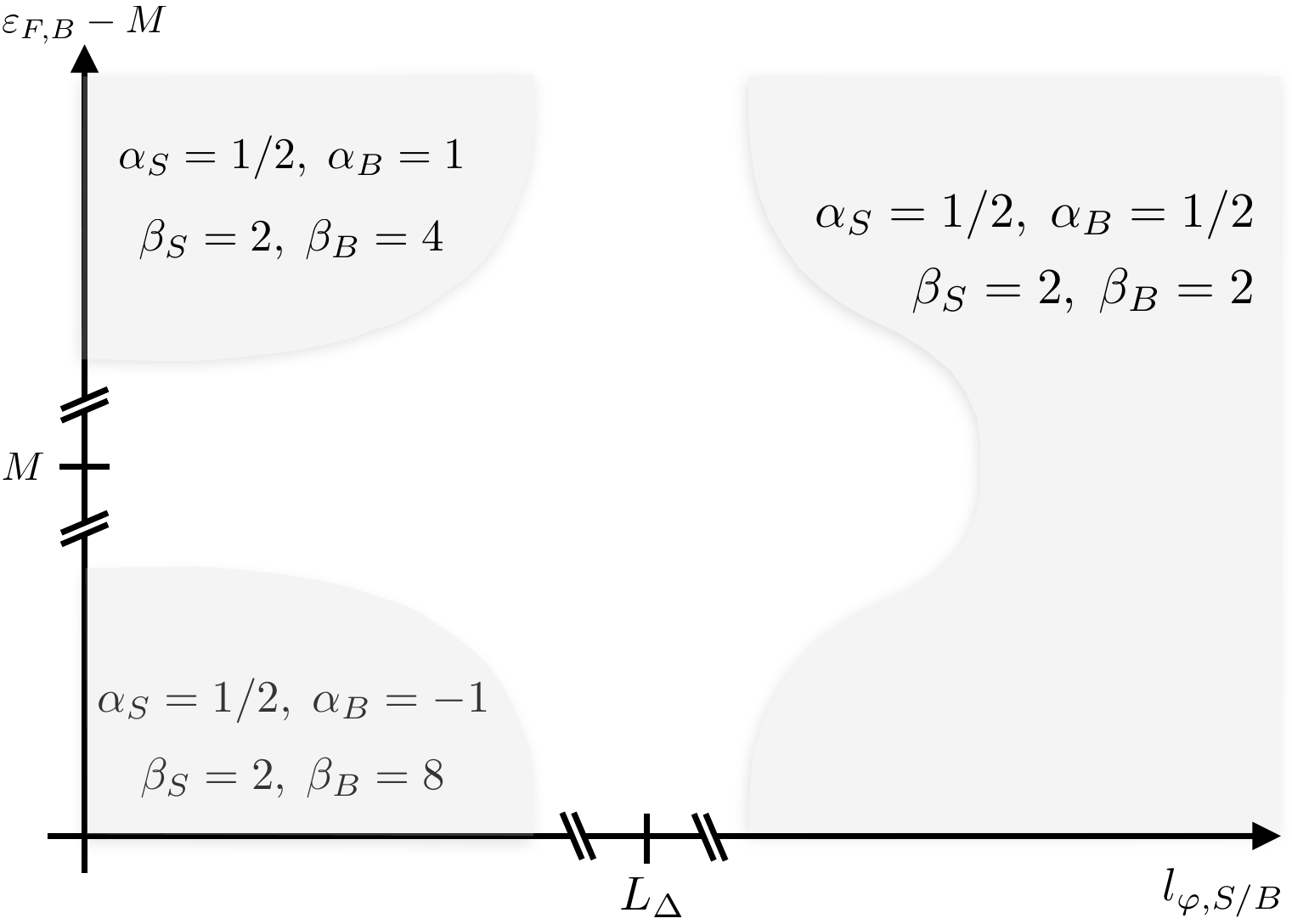}
	\caption{
	Summary of results in the weak-coupling limit. Here $l_{\varphi,S/B}$ refers to the phase decoherence lengths in surface and bulk, respectively,
	and $L_\Delta$ is the length associated to small gaps of soft modes discussed in Tables~II and III. 
 	}
\label{fig:wc}
\end{figure} 

\section*{Acknowledgments}
The authors thank P.~Brouwer, I.~Garate, A.~Levchenko and P.~Silvestrov for useful discussions. H.~V. is a recipient of a DFG-fellowship through the Excellence Initiative by the Graduate School Materials Science in Mainz (GSC 266). T.~M.~acknowledges financial support by Brazilian agencies CNPq and FAPERJ. G.~Schwiete was partially supported by the Alexander von Humboldt Foundation, the College of Arts and Sciences at the University of Alabama, and the National Science Foundation under Grant No. DMR-1742752. We acknowledge the hospitality of the Spin Phenomena Interdisciplinary Center (SPICE), where parts of the work were done. 

\appendix

\section{U-matrices for bulk diffusion modes}

For the sake of notational simplicity, we suppress the bulk index $B$ in this appendix.

In order to calculate the components of the $U$-matrices in Eqs.~\eqref{3D-bulk:diffuson-U-matrix} and X for the diffuson and Cooperon, respectively, it is useful to write the Green's functions \eqref{eq:Green_bulk} in the form
\begin{equation}
	  G_\epsilon^{R/A} (\mathbf{k}) = 
	  \frac{\epsilon^{R/A} + \sum_{\mu=1}^4 \eta_\mu({\bf k}) \Lambda^\mu}{[\epsilon^{R/A} ]^2 - \epsilon_{\bf k}^2},
\label{appendix:3D-bulk-Green-function-so-basis}
\end{equation}
where $\epsilon^{R/A} = \epsilon \pm i\hbar/(2\tau_0)$, $\eta_i(\mathbf{k}) = \hbar v k_i$ for $i \in \{1,2,3\}$ and $\eta_4(\mathbf{k})=M$. The matrices $\Lambda^\mu$ are defined below Eq.~\eqref{3D-bulk:diffuson-U-matrix}. Note that with this notation the clean bulk Hamiltonian reads $H_{3D}({\bf k})=\sum_{\mu=1}^4\eta_\mu({\bf k}) \Lambda^\mu$.

\subsection{Bulk Diffuson}
\label{app:bulk_diffuson}

The coefficients $a$, $b$, $c$ and $d$ appearing in the representation (\ref{3D-bulk:diffuson-U-matrix}) of the matrix $U$ are defined through integrals of the form ($Y\in\{a,b_\mu,c_\mu,d_{\mu\nu}\}$),
\begin{align}
Y=u_0\int (d^3k)\; \frac{\chi_{Y}({\bf k},{\bf q})}{([\epsilon_F^{R} ]^2 - \epsilon_{\bf k}^2)([\epsilon_F^{A} ]^2 - \epsilon_{{\bf k}-{\bf q}}^2)},
\end{align}
and the functions $\chi_Y$ for the different cases are given by
\begin{align}
\chi_a({\bf k},{\bf q})&=\epsilon_F^R \epsilon^A_F,\\
\chi_{b_\mu}({\bf k},{\bf q})&=\epsilon^R_F \eta_\mu (\mathbf{k-q}),\\
\chi_{c_\mu}({\bf k},{\bf q})&=\eta_\mu (\mathbf{k})\epsilon^A_F,\\
\chi_{d_{\mu\nu}}({\bf k},{\bf q})&=\eta_\mu (\mathbf{k}) \eta_\nu (\mathbf{k-q}).
\end{align}

Next, we list the results for the coefficients in the limit $v_Fq\tau_0\ll 1$ (to leading order in $q_i$). To start with,
\begin{align}
a=a_0\left(1-(\ell_0q)^2/3\right),
\label{eq:a}
\end{align}
where $a_0= (2 \left( 1 + M^2/\epsilon^2_F\right))^{-1}$ and $\ell_0=v_F\tau_0$. Further, for $i\in\{1,2\}$, and with $\beta=v_F/v=(1-M^2/\epsilon_F^2)$
\begin{align}
b_i=c_i&=-i(a_0/3)\beta \ell_0 q_i\label{eq:bi},\\
d_{ii}&=(a_0/3)\beta^2 \left[1- \ell_0^2(q^2+2q_i^2)/5\right],\\
d_{i4}=d_{4i}&=-i(a_0/3)\beta(M/\epsilon_F) l_0q_i,
\label{eq:di4}
\end{align}
and further
\begin{align}
b_4=c_4&=(M/\epsilon_F)a \label{eq:b4},\\
d_{12}=d_{21}&=-2(a_0/15)\beta^2\ell_0^2q_1q_2,\\
d_{33}&=(a_0/3)\beta^2\left[1-\ell_0^2 q^2/5\right],\\
d_{44}&=(M^2/\epsilon_F^2)a.\label{eq:d44}
\end{align}
The omitted elements are zero.

\subsection{Bulk Cooperon}
\label{app:bulk_cooperon}
A detailed discussion of the bulk Cooperon has already been presented in Ref.~\onlinecite{Garate12}. In order to make the manuscript self-contained, we display the most important relations below.

The Bethe-Salpeter equation for the Cooperon reads,
\begin{align}
\label{Cooperon-3D-Bethe-Salpeter-so-basis}
	 C^{mn}_{m'n'} (\mathbf{q}) 
	 =
	  u_0
	  \delta_{mn} \delta_{m'n'}
	 +
	  \sum_{l,l'}
	  U^{ml}_{m'l'} (\mathbf{q}) 
	  C^{ln}_{l'n'} (\mathbf{q}), 
\end{align}
with
$U^{ml}_{m'l'} (\mathbf{q}) 
=u_0 \int (d^3 k)
G^{R}_{ml}(\mathbf{k})G^{A}_{m'l'}(-\mathbf{k} + \bold{q})$.

The matrix $U$ appearing in Eq.~\eqref{Cooperon-3D-Bethe-Salpeter-so-basis} can be written as
\begin{align}
\label{3D-bulk:Cooperon-U-matrix}
U^{ml}_{m'l'} ({\bf q})
	 &= 
	 a^c({\bf q}) \, \delta_{ml} \delta_{m'l'}
	 + \sum_\mu b^c_\mu({\bf q}) \delta_{ml} \Lambda^\mu_{l'm'}
\\
	 &\quad
	 + \sum_\mu c^c_\mu({\bf q}) \Lambda^\mu_{ml} \delta_{m'l'} 
	 + \sum_{\mu\nu} d^c_{\mu\nu}({\bf q}) \Lambda^\mu_{ml} \Lambda^\nu_{m'l'}.\nonumber
\end{align}
The coefficients appearing in this formula can be related to those listed in Eqs.~\eqref{eq:a}, \eqref{eq:bi}-\eqref{eq:di4} and \eqref{eq:b4}-\eqref{eq:d44}. One finds $a^c=a$, for $i\in\{1,2\}$ 
\begin{align}
b_i^c&=-c_i^c=-b_i,\\
d_{ii}^c&=d_{ii},\quad d_{i4}^c=-d_{4i}^c=d_{i4},
\end{align}
and further
\begin{align}
b_4^c&=c_4^c=b_4,\\
d^c_{12}&=d_{21}^c=-d_{12},\\
d_{33}^c&=-d_{33},\quad d_{44}^c=d_{44}.
\end{align}
The elements omitted in this list are again equal to zero.

Finally, we write the transformation to the spin-orbit basis with projection onto the conduction band as
\begin{align}
&C^{\beta\beta'}_{\alpha\alpha'}({\bf k}_1,{\bf k}_2,{\bf q})=\sum_{m,m',n'n'}\langle \beta,{\bf k}_1|m\rangle\langle n|\beta',-{\bf k}_2+{\bf q}\rangle\times\nonumber \\
 &\qquad\qquad \times \langle \alpha,-{\bf k}_1+{\bf q}|m'\rangle \langle n'|\alpha',{\bf k}_2\rangle C_{m'n'}^{mn}({\bf q}).
 \label{eq:Cooperon-bulk-transformation}
\end{align}
This is the analog of Eq.~\eqref{3D-bulk-diffuson-transformation} for the bulk diffuson.

\section{Diffusion modes with bulk-surface coupling}
\label{app:BSC}
In this appendix, we discuss the soft diffusion modes, diffusons and Cooperons, for a gated TI thin film, in which the bulk is coupled to a single surface through a tunneling barrier. We focus our attention on the limit in which the tunneling rates $1/\tau_{tB}$ and $1/\tau_{tS}$, Eq.~\eqref{coupling:surface-to-bulk-tunneling-rate}, are much smaller than the disorder scattering rates $1/\tau_{0B}$ and $1/\tau_{0S}$ in \emph{both} bulk and surface. In this limit, one can conveniently use the bulk and surface soft modes in the absence of tunneling as building blocks for the construction of the soft modes of the coupled system. In order to structure the discussion, we follow a two-step procedure based on Ref.~\onlinecite{Garate12}: (i) We define auxiliary diffusion modes, labeled by a tilde sign, which are obtained from the bulk and surface diffusion modes by incorporating the tunneling-induced change of the single-particle propagators. (ii) We obtain the soft modes of the combined bulk-surface system by coupling the auxiliary diffusion modes through correlated particle-hole or particle-particle tunneling events. 

For step (i) we need to account for the change of the surface and bulk Green's functions. These are modified due to the increase of the single-particle scattering rates caused by inter-layer tunneling. The modified Green's functions for bulk and surface, obtained from Eqs.~\eqref{eq:Green_bulk} and \eqref{Gmatrix}, respectively, by the replacement $\tau_{0B/S}^{-1}\rightarrow \tilde{\tau}_{0B/S}^{-1}=\tau_{0B/S}^{-1}+\tau_{tB/S}^{-1}$, will be denoted by $\tilde{G}$. It is convenient to define auxiliary soft modes, both for the surface ($\tilde{\mathcal{C}}$ and $\tilde{\mathcal{D}}$), and bulk ($\tilde{C}$ and $\tilde{D}$), which incorporate the change of the single-particle propagators. As an example, the auxiliary surface diffuson is defined by the Bethe-Salpeter equation,
\begin{align}
\tilde{\mathcal{D}}^{mn}_{m'n'}({\bf q})=u_0\delta_{mn}\delta_{m'n'}+\sum_{l,l'=\uparrow,\downarrow}\tilde{U}^{ml}_{m'l'}({\bf q})\tilde{\mathcal{D}}^{ln}_{l'n'}({\bf q}),
\end{align}
where $\tilde{U}^{ml}_{m'l'}({\bf q})=u_0\int (d^2k)\tilde{G}^{R}_{ml}({\bf k})\tilde{G}^A_{l'm'}({\bf k}
-{\bf q})$. The other auxiliary soft modes are obtained in analogy. The explicit form of these modes is easily deduced from the formulas already provided for the separate bulk and surface soft modes. The most important change is the emergence of an additional mass term of the form $1/\tau_{tB/S}(1+\tau_0/\tau_{tB/S})\approx 1/\tau_{tB/S}$. This mass competes with the phase decoherence rates $1/\tau_{\varphi B/S}$ and, in the case of the almost gapless modes, with the masses $\Delta_{g1}$ and $\Delta_{g2}$. 

The auxiliary soft modes allow formulating the generalized Bethe-Salpeter equations (BSEs) for the soft modes of the coupled system in a compact form. One may distinguish modes for which the initial \emph{and} the final states all live on the surface or all in the bulk from those modes for which surface states eventually transform into bulk states, or vice versa. Only the first kind of modes is relevant for the weak-localization corrections, while the second kind of modes becomes important for the CFs. Here, we will discuss the surface-to-surface modes only, the other modes are obtained in analogy. The BSE for the surface Cooperon mode is shown diagrammatically in Fig~\ref{fig:cc} as an example. 
\begin{figure}[tb]
	\centering
	\includegraphics[scale=0.25]{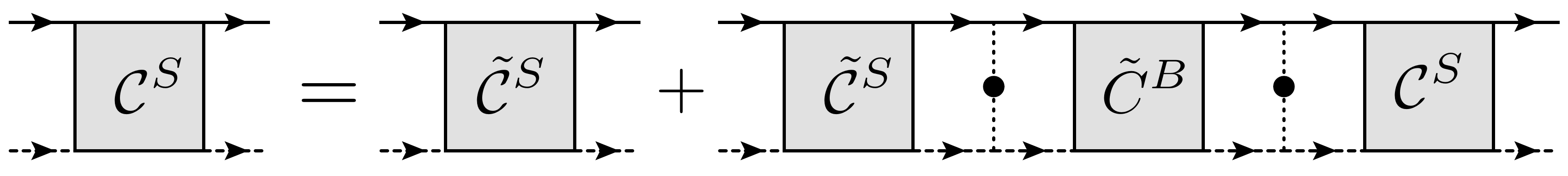}
	\caption{Diagrammatic representation of the Bethe-Salpeter equation for the surface Cooperon in the presence of bulk-surface coupling, Eq.~\eqref{eq:CS}. The dotted line with a dot represents correlated tunneling events.}
\label{fig:cc}
\end{figure}
Written in the eigenbasis of $\tau_z\otimes \sigma_z$ and $\sigma_z$ for bulk and surface, respectively, the BSEs read
\begin{align}
\mathcal{C}^S&=\tilde{\mathcal{C}}^S+\gamma\tilde{\mathcal{C}}^SJ_C^{SB}\tilde{C}^BJ_C^{BS}\mathcal{C}^S\label{eq:C_S},\\
\mathcal{D}^S&=\tilde{\mathcal{D}}^S+\gamma\tilde{\mathcal{D}}^SJ^{SB}_D\tilde{D}^BJ^{BS}_D\mathcal{D}^S\label{eq:DS},
\end{align}
where $\gamma=[(8\pi \nu_B
\tau_{t2})(4\pi \nu_S\tau_{t1})(u_0^Su_0^B)^2]^{-1}$. For the sake of clarity, we introduced explicit labels $S$ and $B$ for surface and bulk quantities and $C$ and $D$ for Cooperon and diffuson related quantities, respectively. The junction matrices $J$ connecting bulk and surface modes read $J_{C/D}^{SB}=\tilde{U}^S_{C/D}T_{C/D}^{SB}\tilde{U}^B_{C/S}$, $J_{C/S}^{BS}= \tilde{U}^{B}_{C/S}T_{C/S}^{BS}U^{S}_{C/S}$. The newly defined matrices $T_{C/D}^{SB/BS}$ reflect the orbital and spin structure of the correlated tunneling events, $(T_C^{SB})^{ab}_{a'b'}=\sum_{\tau,k}\langle a|\tilde{\sigma}^k\otimes \pi_\tau|b\rangle\langle a'|\tilde{\sigma}^k\otimes \pi_\tau|b'\rangle$ and $(T_C^{BS})^{ab}_{a'b'}=\sum_{\tau,k}\langle a|(\tilde{\sigma}^k)^\dagger\otimes \pi^T_\tau| b\rangle\langle a'|(\tilde{\sigma}^k)^\dagger\otimes \pi^T_\tau|b'\rangle$ for the Cooperon and $(T_D^{SB})^{ab}_{a'b'}=\sum_{\tau,k}\langle a|\tilde{\sigma}^k\otimes \pi_\tau|b\rangle\langle b'|(\tilde{\sigma}^k)^\dagger\otimes \pi^T_\tau|a'\rangle$ and $(T_D^{BS})^{ab}_{a'b'}=\sum_{\tau,k}\langle a|(\tilde{\sigma}^k)^\dagger \otimes \pi_{\tau}^T|b\rangle \langle b'|\tilde{\sigma}^k\otimes \pi_\tau|a'\rangle$ for the diffuson. The form of the presented equations implies that two consecutive tunneling rungs are excluded in the ladders. This approximation is justified in the limit of small tunneling rates which we assume.

When writing Eqs.~\eqref{eq:C_S} and \eqref{eq:DS}, the relation between the correlated tunneling matrix elements (cf. Eq.~\eqref {coupling:tunneling-Hamiltonian} with the corresponding correlator for $t_l^\tau$) and the tunneling rates $\tau_{tB/S}^{-1}$ of Eq.~\eqref{coupling:surface-to-bulk-tunneling-rate} has been used. Further, we made use of the fact that the bulk soft modes are effectively two-dimensional, so that all diffusion modes and the matrices $\tilde{U}$ depend on a common two-dimensional momentum.

The junction matrices $J$ connect soft bulk and surface modes. These modes are listed in Tables \ref{fsm} and \ref{asm}. 
They are characterized by their spin structure and, in case of the bulk modes, also by their orbital structure. The matrices $T$ determine allowed couplings between soft modes. For the sake of illustration, we discuss the junctions for the Cooperons in more detail. As one can see from the relations $T_C^{SB}\propto |\chi_s^S\rangle\langle\chi_s^B|\otimes (\langle TT|+\langle BB|)$ and $T_C^{BS}\propto |\chi_s^B\rangle \langle \chi_s^S|\otimes (|TT\rangle+|BB\rangle)$, where $|\chi_s^{S/B}\rangle$ denote spin singlet modes on surface and bulk, respectively, only spin singlet modes couple to each other. Both bulk and surface host the fundamental singlet modes. In addition, in the limit $\varepsilon_{F,B}-M\gg M$, a slightly gapped singlet mode is effective in the bulk. This mode, however, does not couple to the surface singlet mode due to the structure of its orbital part $|\Lambda_2\rangle\propto |TB\rangle+|BT\rangle$. We may therefore conclude that in both limiting cases considered in this paper, $(\varepsilon_{F,B}-M)/M\gg 1$ and $(\varepsilon_{F,B}-M)/M\ll1 $, only the fundamental singlet modes in bulk and surface are coupled to each other.

It will be useful to derive the modified bulk and surface fundamental modes explicitly, using again the Cooperon modes as an example. To this end we use the auxiliary fundamental bulk and surface modes in the form 
\begin{align}
\tilde{\mathcal{C}}^S_{\bf q}&=\frac{u_0^S}{\tau_{0S}}\frac{|\mathcal{C}_0\rangle\langle\mathcal{C}_0|}{D_S{\bf q}^2+\frac{1}{\tau_{tS}}+\frac{1}{\tau_{\varphi S}}}\label{eq:CS},\\
\tilde{\mathcal{C}}^B_{\bf q}&=\frac{u_0^B}{\tau_{0B}}\frac{|C_0\rangle\langle C_0|}{D_{B}{\bf q}^2+\frac{1}{\tau_{tB}}+\frac{1}{\tau_{\varphi B}}}\label{eq:CB},
\end{align}
where $|\mathcal{C}_0\rangle=|\chi_s^S\rangle$ and $|C_0\rangle=|\chi_s^B\rangle\otimes |\Lambda_0\rangle$. Solving the Bethe-Salpeter equations, one obtains
\begin{align}
&C^S_0({\bf q})\equiv \langle \mathcal{C}_0|C^S_{\bf q}|\mathcal{C}_0\rangle=\nonumber\\
&\frac{u_0^S}{\tau_{0S}}\frac{D_B{\bf q}^2+\frac{1}{\tau_{tB}}+\frac{1}{\tau_{\varphi B}}}{(D_S{\bf q}^2+\frac{1}{\tau_{tS}}+\frac{1}{\tau_{\varphi S}})(D_B{\bf q}^2+\frac{1}{\tau_{tB}}+\frac{1}{\tau_{\varphi B}})-\frac{1}{\tau_{tS}\tau_{tB}}},\label{eq:CS1}\\
&C^B_0({\bf q})\equiv\langle C_0|C^B_{\bf q}|C_0\rangle=\nonumber\\
&\frac{u_0^B}{\tau_{0B}}\frac{D_S{\bf q}^2+\frac{1}{\tau_{tS}}+\frac{1}{\tau_{\varphi S}}}{(D_B{\bf q}^2+\frac{1}{\tau_{tB}}+\frac{1}{\tau_{\varphi B}})(D_S{\bf q}^2+\frac{1}{\tau_{tS}}+\frac{1}{\tau_{\varphi S}})-\frac{1}{\tau_{tS}\tau_{tB}}},\label{eq:CB1}
\end{align}
where $D_B=v_F^2\tau_B/3$ is the bulk diffusion coefficient with $v_F=v\sqrt{1-M^2/\varepsilon_{F,B}^2}$, and the surface diffusion coefficient $D_S$ was introduced below Eq.~\eqref{Dsurface}. When writing Eqs.~\eqref{eq:CS}, \eqref{eq:CB}, \eqref{eq:CS1} and \eqref{eq:CB1}, we kept the leading order in the small parameter(s) $\tau_{0S}/\tau_{tS}$ and $\tau_{0B}/\tau_{tB}$ only. We added phenomenological phase decoherence rates $1/\tau_{\varphi S/B}$ for the surface and bulk Cooperons. These expressions can be rewritten in the form given in Eq.~\eqref{eq:CX} of the main text. The expressions for the coupled diffusons $D_0^S$ and $D_0^B$ are found by analogous considerations and take the same form as those stated for $C_0^S$ and $C_0^S$ above.

\section{Conductance fluctuations}
\label{app:UCF}
In this section, we present details on the calculation of the conductance fluctuations.\cite{Lee85,Altshuler85,Altshuler85a,Altshuler86,Lee87} The calculation of CFs in disordered systems has a long history and is well documented.\cite{Lee85,Altshuler86,Akkermans2007} This is why we mainly stress those aspects that are characteristic for the problem at hand. CFs on the surface of a $3d$ topological insulator in the absence of bulk-surface coupling have been studied in Ref.~\onlinecite{Adroguer12}, and our results for this specific case agree with those reported there. 

The conductance fluctuations can be obtained from the fluctuations of the conductivity with the help of the relation $G=\sigma L^{d-2}$, where $L$ is the linear dimension of the sample. We will therefore be concerned with the calculation of the second moment of the conductivity, averaged over impurity configurations. The fluctuations of the conductivity may be decomposed into two contributions - fluctuations of the diffusion coefficient (DCF) and fluctuations of the density of states (DOS).\cite{Altshuler86,Akkermans2007} On a diagrammatic level, both contributions involve the pairing of four trajectories, instead of two as was the case for the WL corrections. To leading order in the small parameter $(k_F l)^{-1} \ll 1$, only diagrams involving products of two Cooperons or two diffusons contribute, see Fig.~\ref{figure:bulk_ucf_collective_diagram}. 
The main building blocks for the calculation, the disorder-averaged Green's functions for the bulk and the surface and the soft Cooperon and diffuson modes for the surface, the bulk, and the coupled system have already been discussed in Sec.~\ref{sec:soft_modes}.

For the DOS contribution the diagrams depicted in Fig.~\ref{DOS_combined} need to be supplemented with their complex-conjugate partner diagrams, in which retarded and advanced Green's functions are interchanged. The diagrams of Fig.~\ref{DOS_combined} give a real contribution, so we can account for the complex-conjugate partners by multiplying the result by a factor 2. Further, both DOS and DCF-type contributions can also be realized with two diffusons. Even though the Cooperon and diffuson matrices differ, in the end both give the same results for the conductance fluctuations. In this appendix, we will discuss the calculation of the conductance fluctuations based on the diagrams depicted in Figs.~\ref{DOS_combined} and \ref{DCF_combined}. According to the arguments given above, we can obtain the final result by multiplying the obtained results by a factor of 4 for the DOS-type diagrams and by a factor of 2 for the DCF-type diagrams.

In the following, we first discuss CFs in bulk and surface separately, and then generalize to the tunneling-coupled system. In all cases we investigate the $x$-component of the conductivity ($xx$-component of the conductivity fluctuations).

\subsection{Bulk}
\label{subsec:bulk}
In this section, DOS and DCF-type fluctuations will be discussed separately for the bulk of the TI. The `bulk' label B will be suppressed in this section since we are dealing with bulk quantities exclusively.
\subsubsection{DOS-type contributions}
As a model example, we discuss the diagram labeled as DOS 1 in Fig.~\ref{DOS_combined}, when the soft modes are Cooperons.

\begin{figure}[h]
	\centering
	\includegraphics[scale=0.45]{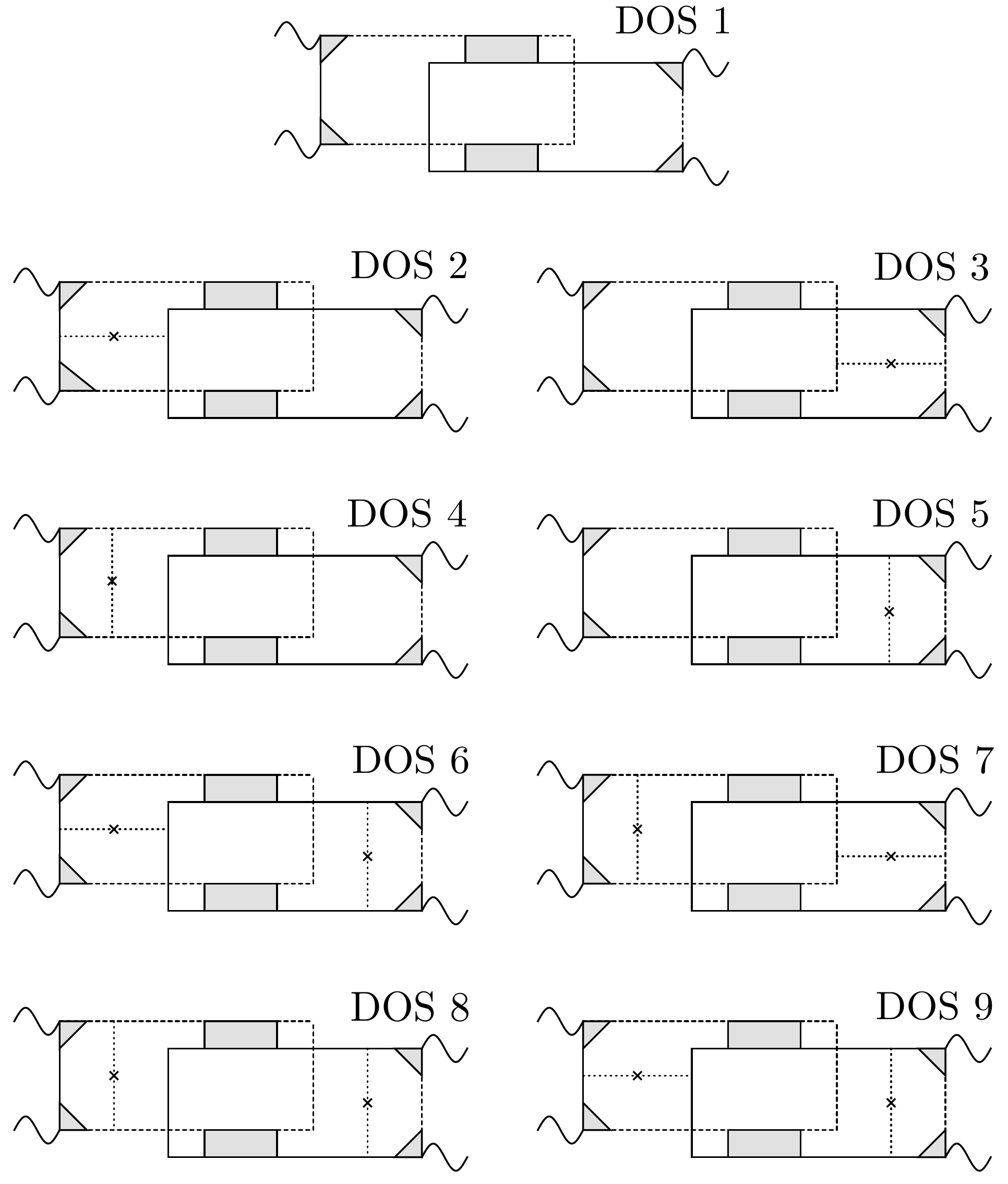}
	\caption{DOS-type diagrams for the calculation of the CFs as discussed in Appendix~\ref{app:UCF}.}
\label{DOS_combined}
\end{figure}

\begin{figure}[h]
	\centering
	\includegraphics[scale=0.37]{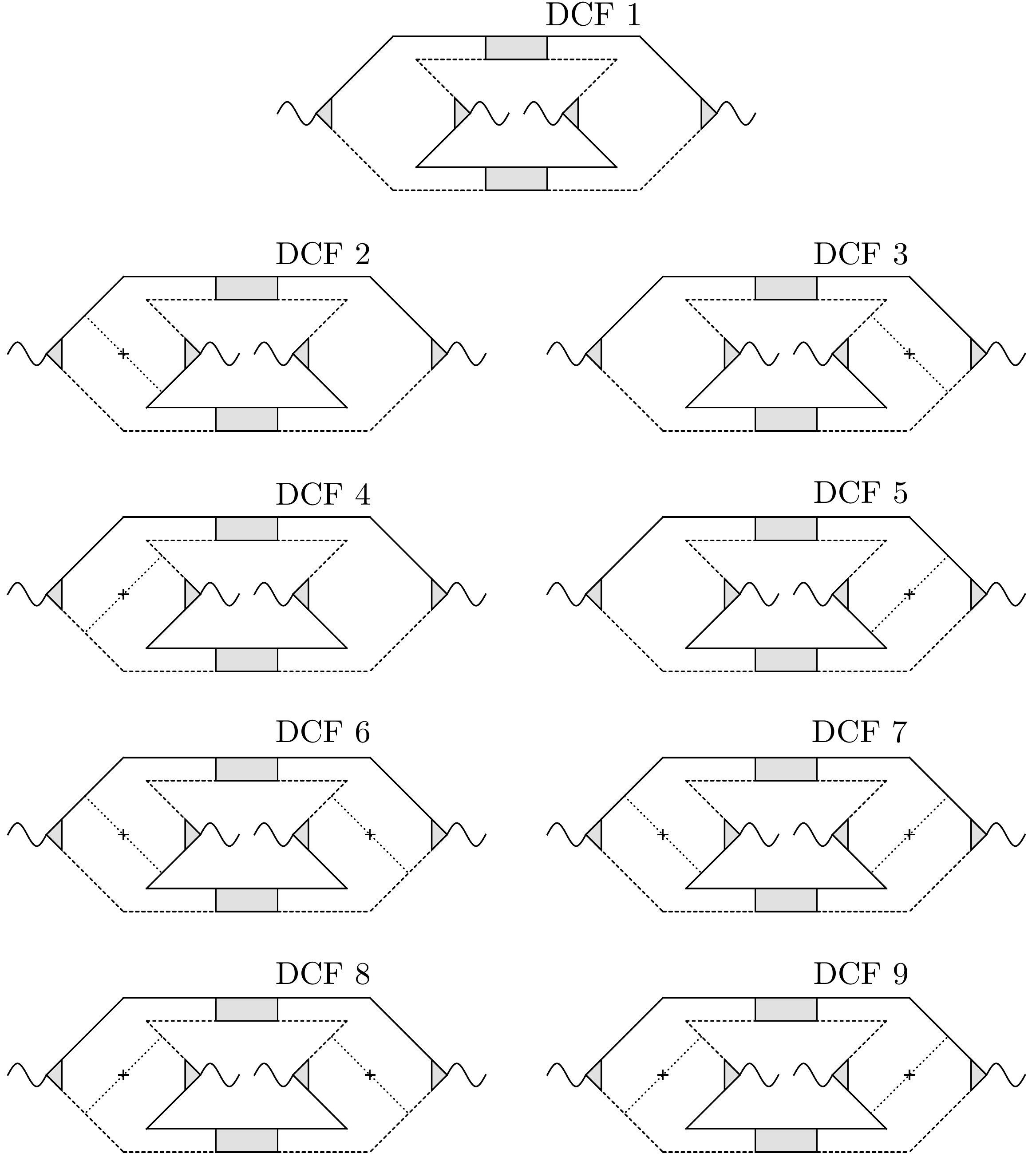}
	\caption{DCF-type diagrams for the calculation of the CFs as discussed in Appendix~\ref{app:UCF}.}
\label{DCF_combined}
\end{figure}
The correction to the conductivity fluctuation originating from this diagram reads as
\begin{align}
&\delta \sigma^2_{\tiny \mbox{DOS\;1, C}}
	=
	\left( \frac{e^2 \hbar}{2 \pi L} \right)^2
	\int_{\mathbf{k}, \mathbf{k}'}
 		  \left[
 		   \tilde{v}^x_{\mathbf{k}}
		    \tilde{v}^x_{\mathbf{k}'}
 		  \right]^2
		   \left[ G^R_\mathbf{k} 	 G^A_\mathbf{k} 	\right]^2\left[ G^R_{\mathbf{k}'}	 G^A_{\mathbf{k}'} 	\right]^2
		   \nonumber \\
		&\times 
		   \frac{1}{W^2} \int_{\mathbf{q}}  \,  
		   \sum_{\substack{\alpha,\alpha' \beta,\beta'=1}}^2C^{\beta'\beta}_{\alpha \alpha'} (\mathbf{k},\mathbf{k}',\mathbf{q}) \,
		   C^{\beta\beta'}_{\alpha'\alpha} (\mathbf{k}',\mathbf{k},\mathbf{q}),
\end{align} 
where it was used that the (renormalized) current vertices $\tilde{v}$ as well as the Green's functions are diagonal in the band indices $\alpha,\beta,$ etc. Important momenta ${\bf k}$ and ${\bf k}'$ under the integral are close to the Fermi surface. This is why the angular integral in the solid angles ${\bf n}_{\bf k}$ and ${\bf n}_{\bf k'}$ may be separated from the integration over the moduli of ${\bf k}$ and ${\bf k}'$. For the purpose of the discussion we define   
\begin{align}
\mathcal{I}({\bf q})=&\sum_{\substack{\alpha,\alpha' \beta,\beta'}}\int \frac{ d{\bf n}_{\bfk}d{\bf n}_{\bfk'}}{(4\pi)^2} \hat{e}_{\bfk}^2\hat{e}_{\bfk'}^2 C^{\beta'\beta}_{\alpha \alpha'} (\mathbf{k}_F,\mathbf{k}'_F,\mathbf{q}) \nonumber
\\ &\times
		   C^{\beta\beta'}_{\alpha'\alpha} (\mathbf{k}'_F,\mathbf{k}_F,\mathbf{q}).
\end{align}
At small momenta, the integral $\mathcal{I}({\bf q})$ is dominated by the soft Cooperon modes discussed in Sec.~\ref{sec:soft_modes}. In the limits of large and small Fermi energies one obtains
\begin{align}
\mathcal{I}(\bfq)=\left\{\begin{array}{cc} \frac{1}{36}\left[C^2_0({\bf q})+C_s^2({\bf q})\right],& \varepsilon_F-M\gg M,\\
\frac{1}{9}\left[C_0^2({\bf q})+\sum_{m} C_{t,m}^2({\bf q})\right],& \varepsilon_F-M\ll M.\end{array}\right. 
\end{align}
Here, we employed the fundamental Cooperon mode $C_0({\bf q})=\langle C_0|C_{\bf q}|C_0\rangle$ as well as the soft modes in limits of large and small Fermi energies, $C_s({\bf q})=\langle C_s|C_{\bf q}|C_s\rangle$ and $C_{t,m}({\bf q})=\langle C_{t,m}|C_{\bf q}|C_{t,m}\rangle$ for $m\in\{1,2,3\}$, respectively. With the help of this result, one finds
\begin{align}
&\delta \sigma^2_{\tiny \mbox{DOS\;1, C}}
	=\left\{\begin{array}{cc} l^{\tiny\mbox{DOS}}_1\\s^{\tiny \mbox{DOS}}_1\end{array}\right\}
	\left( \frac{e^2}{h L} \right)^2\left(\frac{\tau}{\tau_0}\right)^2\frac{\mathcal{\mathcal{I}_S}}{W^2},
	\label{eq:s2dos} 
\end{align}
where $l^{\tiny\mbox{DOS}}_1=s^{\tiny\mbox{DOS}}_1=4$. $\mathcal{I}_S$ stands for the momentum integral over the soft diffusion modes,
\begin{align}	
	&\mathcal{I}_S= \int_\bfq \left\{\begin{array}{cc}  \frac{1}{\left({\bf q}^2+l_{\varphi}^{-2}\right)^2}+\frac{1}{\left({\bf q}^2+l_{g2}^{-2}+l_{\varphi}^{-2}\right)^2},
	&\varepsilon_{F}-M\gg M,\\
 \frac{1}{\left({\bf q}^2+l_{\varphi}^{-2}\right)^2}+\frac{3}{\left({\bf q}^2+l_{g1}^{-2}+l_{\varphi}^{-2}\right)^2},&\varepsilon_{F}-M\ll M,\end{array}\right.
\label{eq:IS}
\end{align}
where $l^{-2}_{\varphi}=1/D\tau_\varphi$ and $l^{-2}_{gi}=\Delta_{gi}/{D}$.

The contributions from diagrams DOS 2-DOS 9 can also be written in the form of Eq.~\ref{eq:s2dos} with varying coefficients $l^{\tiny\mbox{DOS}}_i$ and $s^{\tiny\mbox{DOS}}_i$. They are listed in table \ref{table:UCF}. These diagrams contain additional disorder lines as compared to DOS 1. For the calculation, one should remember that the matrix elements of the disorder potential are nontrivial in the eigenbasis. 

From table \ref{table:UCF}, we read off $\sum_{i=1}^9l_{i}^{\tiny\mbox{DOS}}=4/9$ and $\sum_{i=1}^9s_{i}^{\tiny\mbox{DOS}}=1$. Further, taking into account $\tau\approx 3\tau_0/2$ for $\varepsilon_F-M/\gg M$ and $\tau\approx \tau_0$ for $\varepsilon_F-M\ll M$, we find
\begin{align}
&\delta \sigma^2_{\tiny\mbox{DOS,C}}=\sum_{i=1}^9\sigma^2_{\tiny\mbox{DOS\;i,C}}=\left( \frac{e^2 }{h L} \right)^2\frac{\mathcal{\mathcal{I}_S}}{W^2}. 
\end{align}

\begin{table}[t!]
\begin{tabular}{l|c|c|c|c}
\hline
\hline
 & bare& \hspace{0.3cm} one imp.ln.  & two  imp.lns.&total
  \\
  &i=1&i=2-5&i=6-9&\\
\hline 
{\bf Bulk}&&&&\\
{\bf $\varepsilon_F-M\gg M$}&&&&\\
 $l^{\tiny\mbox{DOS}}_i$& $4$ & $-\tfrac{2}{3}-\tfrac{2}{3}-2-2$  
 & $\tfrac{1}{3}+\tfrac{1}{3}+1+\tfrac{1}{9}$&$\frac{4}{9}$
\\
 $l^{\tiny\mbox{DCF}}_i$ & $ 4$ 
 & $\;-\tfrac{2}{3}-\tfrac{2}{3}-\tfrac{2}{3}-\tfrac{2}{3}\;$ 
 & $\tfrac{1}{9}+\tfrac{1}{9}+\tfrac{1}{9}+\tfrac{1}{9}$  &$\frac{16}{9}$
 \\
{\bf $\varepsilon_F-M\ll M$}&&&&\\
 $s^{\tiny\mbox{DOS}}_i$ & $4$& $0+0-2-2$  & $0+0+1+0$&$1$
\\
 $s^{\tiny\mbox{DCF}}_i$ & $4$& $0+0+0+0$  & $0+0+0+0$  & $4$
 \\
\hline 
\hline
{\bf Surface}&&&&\\
 $c^{\tiny\mbox{DOS}}_i$ & $4$&  $-1-1-2-2$ 
 & $\tfrac{1}{2}+ \tfrac{1}{2}+1+\tfrac{1}{4}$&$\frac{1}{4}$
\\
 $c^{\tiny\mbox{DCF}}_i$ & $4$&  $-1-1-1-1$ 
 &  $\;\tfrac{1}{4}+ \tfrac{1}{4}+ \tfrac{1}{4}+\tfrac{1}{4}\;$&$1$
 \\
\hline
\hline
{\bf Coupling}&&&&\\
{\bf $\varepsilon_F-M\gg M$}&&&&\\
$\tilde{l}^{\tiny\mbox{DOS}}_i$ &$8$& $-\frac{5}{3}-\frac{5}{3}-4-4$ & $\frac{5}{6}+\frac{5}{6}+2+\frac{1}{3}$&$\frac{2}{3}$\\
$\tilde{l}^{\tiny\mbox{DCF}}_i$ & $8$&$-\frac{5}{3}-\frac{5}{3}-\frac{5}{3}-\frac{5}{3}$&$\frac{1}{3}+\frac{1}{3}+\frac{1}{3}+\frac{1}{3}$&$\frac{8}{3}$\\
{\bf $\varepsilon_F-M\ll M$}&&&&\\
$\tilde{s}^{\tiny\mbox{DOS}}_i$ & $8$& $-1-1-4-4$& $\frac{1}{2}+\frac{1}{2}+2+0$&$1$\\
$\tilde{s}^{\tiny\mbox{DCF}}_i$ &$8$&$-1-1-1-1$&$0+0+0+0$&$4$\\
\hline 
\hline
  \end{tabular}
\caption{Numerical factors of the nine diagrams contributing to fluctuations 
of the density of states, $\delta\nu$, and the diffusion constant, $\delta D$. 
`Bare' here refers to diagrams without single impurity lines, `one imp.ln.' and `two  imp.lns.'
indicates the corresponding number of the latter.
The first two (third and fourth) lines give the bulk contribution in the limit of large (small) Fermi energies. 
The next two lines summarize surface contributions, the contributions listed in the last four lines arise in the presence of a finite bulk surface coupling only. The coefficients listed here should be introduced into Eqs.~\eqref{eq:s2dos}, \eqref{eq:s2dcf}, \eqref{ds2surface}, or \eqref{dgbdgs} in order to find the contributions of the respective diagrams.
}
\label{table:UCF}
\end{table}

\subsubsection{DCF-type contribution}
Here, we discuss the diagram labeled DCF 1 in Fig.~\ref{DCF_combined} with soft Cooperon modes as an example,
\begin{align}
&\delta \sigma^2_{DCF 1,C}=\left(\frac{e^2\hbar}{2\pi L}\right)^2\sum_{\alpha,\alpha',\beta,\beta'=1}^2\int_{\bfk,\bfk'}\tilde{v}_{\beta'}^x({\bf k})\tilde{v}_{\alpha}^x(-{\bf k})\nonumber\\
&\qquad \times \tilde{v}_{\alpha'}^x({\bf k'})\tilde{v}_\beta^x(-{\bf k'})\left[G^R_{\bfk}G^A_{\bfk}\right]^2\left[G_{\bfk'}^RG_{\bfk'}^A\right]^2\nonumber\\
&\qquad \times \frac{1}{W^2}\int_{\bfq}C_{\alpha\alpha'}^{\beta'\beta}({\bf k},{\bf k'},{\bf q})C_{\beta,\beta'}^{\alpha'\alpha}({\bf k'},{\bf k},{\bf q}).
\end{align}
Proceeding as for the DOS contribution, one finds
\begin{align}
&\delta \sigma^2_{\tiny \mbox{DCF\;1, C}}
	=\left\{\begin{array}{cc} l^{\tiny\mbox{DCF}}_1\\s^{\tiny \mbox{DCF}}_1\end{array}\right\}
	\left( \frac{e^2 }{h L} \right)^2\left(\frac{\tau}{\tau_0}\right)^2\frac{\mathcal{\mathcal{I}_S}}{W^2},
	\label{eq:s2dcf} 
\end{align}
with $l^{\tiny\mbox{DCF}}_{1}=s^{\tiny \mbox{DCF}}_1=4$. The integral $\mathcal{I}_S$ was defined in Eq.~\eqref{eq:IS}. The overall structure of this result coincides with that of the DOS diagrams. The contributions from diagrams DCF 2--DCF 9 can also be written in the form of Eq.~\eqref{eq:s2dcf} with varying coefficients $l^{\tiny\mbox{DCF}}_i$ and $s^{\tiny\mbox{DCF}}_i$. They are listed in table \ref{table:UCF}. Adding the contributions of the nine diagrams, we obtain
\begin{align}
&\delta \sigma^2_{\tiny \mbox{DCF, C}}=\sum_{i=1}^9\sigma^2_{\tiny \mbox{DCF\;i, C}}=4
	\left( \frac{e^2}{h L} \right)^2\frac{\mathcal{\mathcal{I}_S}}{W^2}.
\end{align}

\subsubsection{Result--CFs bulk}
As mentioned earlier, the total result including all relevant diagrams with Cooperons and the analogous diagrams for diffusons is obtained as a weighted sum, 
\begin{align}
\delta\sigma^2=4\times \delta\sigma^2_{\tiny \mbox{DOS, C}}+2\times \delta \sigma^2_{\tiny \mbox{DCF, C}}=12\left( \frac{e^2}{h L} \right)^2\frac{\mathcal{\mathcal{I}_S}}{W^2} .\nonumber 
\end{align}
Using $G=\sigma W$, we can formulate the final result for the conductance fluctuations of the bulk, 
\begin{align}
\delta G^2=&12\left( \frac{e^2 }{h L} \right)^2\times \\
&\times \int_\bfq \left\{\begin{array}{cc}  \frac{1}{\left({\bf q}^2+l_{\varphi}^{-2}\right)^2}+\frac{1}{\left({\bf q}^2+l_{g2}^{-2}+l_{\varphi}^{-2}\right)^2}, 
&\varepsilon_F-M\gg M,\\
 \frac{1}{\left({\bf q}^2+l_{\varphi}^{-2}\right)^2}+\frac{3}{\left({\bf q}^2+l_{g1}^{-2}+l_{\varphi}^{-2}\right)^2}, 
 &\varepsilon_F-M\ll M. \end{array}\right.\nonumber
\end{align} 
For the sake of simplicity, this formula assumes the same phase coherence time for bulk Cooperons and bulk diffusons.

\subsection{Surface}
\label{subsec:surface}
In this section, DOS and DCF-type fluctuations will be discussed separately for the surface of the TI. The `surface' label S will be suppressed in this section since we are dealing with surface quantities exclusively.

The calculation for the surface proceeds in analogy to the bulk. For example, the first DOS diagram reads as 
\begin{align}
\delta \sigma^2_{\tiny \mbox{DOS\;1, C}}
	=&
	\left( \frac{e^2 \hbar}{2 \pi L} \right)^2
	\int_{\mathbf{k}, \mathbf{k}'}
 		  \left[
 		   \tilde{v}^x_{\mathbf{k}}
		    \tilde{v}^x_{\mathbf{k}'}
 		  \right]^2
		   \left[ G^R_\mathbf{k} 	 G^A_\mathbf{k} 	\right]^2\left[ G^R_{\mathbf{k}'}	 G^A_{\mathbf{k}'} 	\right]^2
		   \nonumber \\
		&\times 
		   \int_{\mathbf{q}}  \,  
		\mathcal{C}(\mathbf{k},\mathbf{k}',\mathbf{q}) \,
		   \mathcal{C}(\mathbf{k}',\mathbf{k},\mathbf{q}),
\end{align}
which can be simplified by integration in fast momenta ${\bf k}$ and ${\bf k'}$. The results for the DOS and DCF-type diagrams can be summarized in the form
\begin{align}
\left\{\begin{array}{cc}\delta \sigma^2_{\tiny \mbox{DOS\;i,C}}\\\delta \sigma^2_{\tiny \mbox{DCF\;i,C}}\end{array}\right\}=&\left\{\begin{array}{cc} c_i^{\tiny \mbox{DOS}}\\c_i^{\tiny \mbox{DCF}}\end{array}\right\}\left(\frac{e^2}{hL}\right)^2\left(\frac{\tau}{\tau_0}\right)^2\nonumber \\
&\times \int_{\bfq} \frac{1}{(\bfq^2+l_\varphi^{-2})^2}.\label{ds2surface}
\end{align}
The coefficients $c_i$ are listed in table \ref{table:UCF}. Using $\sum_{i=1}^9c_i^{\tiny \mbox{DOS}}=1/4$ and $\sum_{i=1}^9c_i^{\tiny \mbox{DCF}}=1$ as well as the relation $\tau=2\tau_0$ results in
\begin{align}
\left\{\begin{array}{ccc}\delta \sigma^2_{\tiny \mbox{DOS,C}}\\\delta \sigma^2_{\tiny \mbox{DCF,C}}\\ \end{array}\right\}=&\left\{\begin{array}{ccc}$1$\\$4$\end{array}\right\}\left(\frac{e^2}{hL}\right)^2\int_\bfq \frac{1}{({\bfq^2+l_\varphi^{-2}})^2}.
\end{align}
We are now ready to compute the average second moment of the conductivity on the surface as $\delta\sigma^2=4\times \delta\sigma^2_{\tiny \mbox{DOS, C}}+2\times \delta \sigma^2_{\tiny \mbox{DCF, C}}$, which immediately leads us to 
\begin{align}
\delta G^2&=12\left(\frac{e^2}{hL}\right)^2\int_\bfq \frac{1}{({\bfq^2+l_\varphi^{-2}})^2}.
\end{align}
This result includes contributions from Cooperon modes as well as from diffuson modes.

\subsection{Coupled system}

For the coupled system, we are interested in the fluctuations of the total conductance of the system,
\begin{align}
\delta G^2=(\delta G_B+\delta G_S)^2.
\label{eq:G2coupled}
\end{align}
Averaging with respect to disorder and random tunneling is implied. Cross correlations between bulk and surface appear because the tunneling coupling activates diffusion processes connecting bulk and surface. As a consequence, in the diagrammatic language we may now attribute either of the two electronic loops to the bulk or to the surface. 

The results for $\delta G_{B}^2$ and $\delta G_S^2$ on the right hand side of relation \eqref{eq:G2coupled} may be inferred from those in the absence of tunneling discussed above, with two differences. 
(i) The soft modes $C_{t,m}$ and $C_s$ in the bulk acquire additional damping due to the tunneling process, compare Eqs.~\eqref{eq:Ctx}--\eqref{eq:Ds}. 
(ii) The fundamental diffusion modes $C^S_0$ and $\mathcal{C}^B_0$ now take the form introduced in Eq.~\eqref{eq:CX}. 

For the surface term, only (ii) is important, and one finds 
\begin{align}
\delta G_S^2=6\left(\frac{e^2}{hL}\right)^2\sum_{\alpha=C,D}\int_\bfq \left[\frac{A_{S\alpha}}{\bfq^2+q^2_{a\alpha}}+\frac{B_{S,\alpha}}{\bfq^2+q_{b\alpha}^2}\right]^2,
\end{align}
where $A_S$ and $B_S$, $q_a$ and $q_b$ have been defined in and below Eq.~\eqref{eq:qaqb}, and, for the sake of clarity, we distinguish the expressions for Cooperons and diffusons by the label $\alpha$. 

For the bulk, we need to take into account the additional damping of non-fundamental soft modes, point (i), to obtain 
\begin{align}
\delta G_B^2=&6\left(\frac{e^2}{hL}\right)^2\sum_{\alpha=C,D}\int_\bfq \left(\left[\frac{A_{B\alpha}}{\bfq^2+q^2_{a\alpha}}+\frac{B_{B,\alpha}}{\bfq^2+q_{b\alpha}^2}\right]^2\right.\nonumber\\
&\left.+\left\{\begin{array}{cc} \frac{1}{(\bfq^2+l_{g2}^{-2}+l_{B\alpha}^{-2})^2}&\\\frac{3}{(\bfq^2+l_{g1}^{-2}+l_{B\alpha}^{-2})^2}\end{array}\right\}\right),
\end{align}
where the upper (lower) line corresponds to large (small) Fermi energies. We wrote $l_{gi}^{-2}=\Delta_{gi}/D_B$ and $l_B$ has been defined below Eq.~\eqref{eq:qaqb}.

The term $2\delta G_B\delta G_S$ on the right hand side of Eq.~\eqref{eq:G2coupled} requires an additional calculation. First, let us note that we need to know the off-diagonal Cooperons $C^{SB}_\bfq=|\mathcal{C}_0\rangle C_0^{SB}(\bfq)\langle C_0|$ and $C^{BS}_\bfq=|C_0\rangle C_0^{BS}(\bfq) \langle \mathcal{C}_0|$ connecting bulk and surface states. A diagrammatic representation of the equation for $C^{BS}_\bfq$ is displayed in Fig.~\ref{fig:cbs} as an example. 
\begin{figure}[tb]
	\centering
	\includegraphics[scale=0.4]{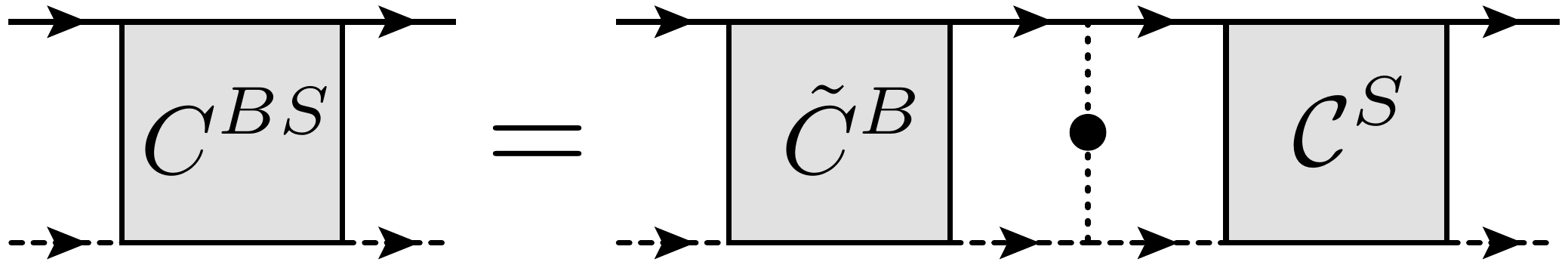}
	\caption{Diagrammatic representation for the Cooperon mode $C^{BS}$ connecting surface and bulk.}
\label{fig:cbs}
\end{figure}
It may be solved by the method described in Appendix~\ref{app:BSC}. In the diagrams always combination $C^{SB}_0(\bfq)C^{BS}_0(\bfq)$ appears. Following similar arguments as those presented in Appendix~\ref{app:BSC} for the coupled system one may now find $C^{SB}_0(\bfq)$ and $C^{BS}_0(\bfq)$. 
We quote here the result for the product,
\begin{align}
C^{SB}_0(\bfq)C^{BS}_0(\bfq)&=\frac{1}{\tau_{tB}\tau_{tS}}\frac{u_0^Bu_0^S}{\tau_{0B}\tau_{0S}}\frac{1}{(D_BD_S)^2}\times\nonumber \\
&\times \left[\frac{C}{\bfq^2+q_a^2}-\frac{C}{\bfq^2+q_b^2}\right]^2,
\end{align}
which appears in all diagrams in this combination. We defined $C=1/(q_b^2-q_a^2)$. An expression of the same form is found for the diffusons.

We use the diagram DOS 1 as an example again. Here, we are interested in the terms where the left velocity vertices are on the surface and the right velocity in the bulk, or vice versa. Adding the two contributions, we get
\begin{align}
&(2\delta G_B\delta G_S)_{\tiny \mbox{DOS}\;1,C}=2\left(\frac{e^2\hbar}{2\pi L}\right)^2\sum_{\alpha\beta'}\int_{\bfk\bfk'}[\tilde{v}_{B,\bfk}^x\tilde{v}_{S,\bfk'}^x]^2\nonumber\\
&\times [G^R_{B,\bfk}G^A_{B,\bfk}]^2
[G^R_{S,\bfk'}G^A_{S,\bfk'}]^2\int_{\bfq} (C_0^{BS})_{\alpha}^{\beta'}({\bfk},{\bfk}',\bfq)\nonumber\\
&\times (C_0^{SB})^{\beta'}_{\alpha}(\bfk',\bfk,{\bfq}),
\end{align}
with
\begin{align}
(C_0^{BS})_{\alpha}^{\beta'}({\bfk},{\bfk}',\bfq)&=\langle\beta',\bfk|\otimes \langle \alpha ,-\bfk|C_\bfq^{BS}|-\bfk'\rangle\otimes |\bfk'\rangle,
\end{align}
compare to Eqs.~\eqref{surface:cooperon-transformation} and \eqref{eq:Cooperon-bulk-transformation}. $(C_0^{SB})^{\beta'}_{\alpha}(\bfk',\bfk,{\bfq})$ is defined in analogy.

After manipulations closely resembling to those presented in Secs.~\ref{subsec:bulk} and \ref{subsec:surface}, the result can be written as
\begin{align}
&(2\delta G_B\delta G_S)_{\tiny \mbox{DOS}\;1,C}=\left\{\begin{array}{cc} \tilde{l}_1^{\tiny \mbox{DOS}}\\\tilde{s}_1^{\tiny \mbox{DOS}}\end{array}\right\}\left(\frac{e^2}{hL}\right)^2\left(\frac{\tau_B}{\tau_{0B}}\right)\left(\frac{\tau_S}{\tau_{0S}}\right)\times\nonumber \\
&\times\int_\bfq\frac{1}{l_{tS}l_{tB}}\left[\frac{C}{\bfq^2+q_a^2}-\frac{C}{\bfq^2+q_b^2}\right]^2,\label{dgbdgs}
\end{align}
with $\tilde{l}_1^{\tiny \mbox{DOS}}=8$ and $\tilde{s}_1^{\tiny \mbox{DOS}}=4$. In a similar way, one can proceed for the remaining eight diagrams of the DOS type and for the DCF-type diagrams. The corresponding coefficients are listed in Table \ref{table:UCF}. Accounting for sum of coefficients quoted in the table and the corresponding ratios of scattering times in the limits of large and small Fermi energies, we obtain
\begin{align}
&(2\delta G_B\delta G_S)_{\tiny \mbox{DOS},C}
=\sum_{i=1}^9(2\delta G_B\delta G_S)_{\tiny \mbox{DOS},C i}\nonumber\\
&=2\left(\frac{e^2}{hL}\right)^2\int_\bfq\frac{1}{l_{tS}l_{tB}}\left[\frac{C}{\bfq^2+q_a^2}-\frac{C}{\bfq^2+q_b^2}\right]^2,
\end{align}
in both limits of large and small Fermi energies. 
For the DCF terms, we find in a similar way
\begin{align}
&2(\delta G_B\delta G_S)_{\tiny \mbox{DCF},C}=\sum_{i=1}^9(2\delta G_B\delta G_S)_{\tiny \mbox{DCF},C i}\nonumber\\
&=8\left(\frac{e^2}{hL}\right)^2\frac{1}{l_{tS}l_{tB}}\int_\bfq\left[\frac{C}{\bfq^2+q_a^2}-\frac{C}{\bfq^2+q_b^2}\right]^2.
\end{align}
We are now in a position to add all relevant contributions,
\begin{align}
&(2\delta G_B\delta G_S)\\
=&\sum_{\alpha=C,D}\left(2\times (2\delta G_B\delta G_S)_{\tiny \mbox{DOS},\alpha}+(2\delta G_B\delta G_S)_{\tiny \mbox{DCF},\alpha}\right),\nonumber
\end{align}
and obtain
\begin{align}
&2\overline{\delta G_B\delta G_S}\\
=&12\sum_{\alpha=C,D}\left(\frac{e^2}{hL}\right)^2\frac{1}{l_{tS}l_{tB}}\int_\bfq\left[\frac{C_\alpha}{\bfq^2+q_{a\alpha}^2}-\frac{C_\alpha}{\bfq^2+q_{b\alpha}^2}\right]^2.\nonumber
\end{align}

Finally, we need to combine all terms on the right hand side of Eq.~\eqref{eq:G2coupled}. In order to simplify the result, the following relations are useful,
\begin{align}
&A^2_{B\alpha}+2C_\alpha^2/(l_{tS}l_{tB})^2+A^2_{S\alpha}=1,\\
&B^2_{B\alpha}+2C_\alpha^2/(l_{tS}l_{tB})^2+B^2_{S\alpha}=1,\\
&A_{S\alpha}B_{S\alpha}=A_{B\alpha}B_{B\alpha}=C_\alpha^2/(l_{tS}l_{tB})^2.
\end{align}
The final result for the conductance fluctuations of the system with bulk-surface coupling takes a simple form, 
\begin{align}
\delta G^2=&6\left(\frac{e^2}{hL}\right)^2\sum_{\alpha=C,D}\int_\bfq \left(\frac{1}{(\bfq^2+q^2_{a\alpha})^2}+\frac{1}{(\bfq^2+q_{b\alpha}^2)^2}\right.\nonumber\\
&\left.+\left\{\begin{array}{cc} \frac{1}{(\bfq^2+l_{g2}^{-2}+l_{B\alpha}^{-2})^2}\\\frac{3}{(\bfq^2+l_{g1}^{-2}+l_{B\alpha}^{-2})^2}\end{array}\right\}\right),\label{eq:G2finalapp}
\end{align}
where the upper (lower) line is applicable for large (small) Fermi energies, $l_{gi}$ has been defined below Eq.~\eqref{eq:IS} and $l_B$ below Eq.~\eqref{eq:qaqb}

It is instructive to recast the result in the form of Eq.~\eqref{eq2a} in the main text.
In Eq.~\eqref{eq2a}, $L_{\Delta_\alpha}$ is the length scale corresponding to the gap of the soft mode $\alpha$ (obtained explicitly by comparison with \eqref{eq:G2finalapp}). Having in mind a finite system size, the momentum integral in Eq.~\eqref{eq2a} has been converted into a sum over discrete momenta. The specific form of the sum is determined by the boundary conditions, as will be discussed next.

\subsection{Final result}
\label{subsec:final}
So far, the quantization of momenta ${\bf q}$ due to the finite system size was not accounted for explicitly. One can model ideal leads coupled at $x=0$ and $x=L$ by absorbing walls (Dirichlet boundary conditions), while the boundary of the sample at $y=0$ and $y=L$ may be modeled by reflecting walls (van Neumann boundary condition). This reasoning corresponds to the replacement of the integral in ${\bf q}$ by a sum over mode numbers $n_x,n_y$ with $n_x=0$  excluded.~\cite{Akkermans2007} 

Introducing the dimensionless function $\mathcal{F}$ as 
\begin{align}
\mathcal{F}(x)=\frac{6}{\pi^4}\sum_{n_x=1}^\infty\sum_{n_y=0}^\infty \frac{1}{(n_x^2+n_y^2+x^2/\pi^2)^2},
\label{eq:Fsum}
\end{align}
we may finally cast the result in the form of Eq.~\eqref{eq:genresult} in the main text, 
where, as in Eq.~\eqref{eq2a}, $\alpha$ runs over all soft diffusion modes.

\end{document}